%% file: sample701.tex
\pdfoutput=1
\documentclass[twocolumn]{aastex701}

\usepackage{lineno}
\linenumbers
\usepackage[T1]{fontenc}
\usepackage{placeins}
\usepackage{float}
\usepackage{graphicx}	
\usepackage{amsmath}	
\usepackage{subfigure}
\usepackage{longtable}
\usepackage{supertabular}
\usepackage{paracol}
\usepackage{tikz}
\usetikzlibrary{calc}
\usepackage{appendix}

\begin{document}

\title{SN 2019cqc: A Hydrogen-rich Superluminous Supernova Showing Electron-Scattering Signatures}

\author[0009-0009-3168-8868]{Dinesh Hebbar}
\affiliation{Manipal Centre for Natural Sciences, Manipal Academy of Higher Education, Manipal 576104, Karnataka, India}
\email{dineshgh82@gmail.com}  

\author[0000-0002-9711-6207]{Rupak Roy}
\affiliation{Institute of Astronomy Space and Earth Science (IASES), P 177, CIT Road, Scheme 7m, Kolkata-700054, West Bengal, India}
\email{rupakroy1980@gmail.com}

\author[0000-0002-2555-3192]{Matt Nicholl}
\affiliation{Astrophysics Research Centre, School of Mathematics and Physics, Queen's University Belfast, Belfast BT7 1NN, UK}
\email{}

\author[orcid=0000-0003-1546-6615,sname='Sollerman']{Jesper Sollerman}
\affiliation{Department of Astronomy, The Oskar Klein Center, Stockholm University, AlbaNova University Center, SE 106 91 Stockholm, Sweden}
\email[]{}

\author[orcid=0000-0003-0227-3451,sname='Anderson']{Joseph P Anderson}
\affiliation{European Southern Observatory, Alonso de Córdova 3107, Vitacura, Casilla 19001, Santiago, Chile}
\email[]{}

\author[orcid=0000-0002-1650-1518 ,sname='Gromadzki']{Mariusz Gromadzki}
\affiliation{Astronomical Observatory, University of Warsaw, Al. Ujazdowskie 4, 00-478 Warszawa, Poland}
\email[]{}

\author[orcid=0000-0003-4253-656X,sname='Howell']{D. Andrew Howell}
\affiliation{Las Cumbres Observatory, 6740 Cortona Drive, Suite 102, Goleta, CA 93117-5575, USA}
\affiliation{Department of Physics, University of California, Santa Barbara, CA 93106-9530, USA}
\email[]{}

\author[orcid=0000-0002-5477-0217,sname='Kangas']{Tuomas Kangas}
\affiliation{Finnish Centre for Astronomy with ESO (FINCA), FI-20014 University of Turku, Finland}
\affiliation{Department of Physics and Astronomy, FI-20014, University of Turku, Finland}
\email[]{}

\author[orcid=0000-0001-7497-2994,sname='Mattila']{Seppo Mattila}
\affiliation{Department of Physics and Astronomy, FI-20014, University of Turku, Finland}
\affiliation{School of Sciences, European University Cyprus, Diogenes Street, Engomi, 1516 Nicosia, Cyprus.}
\email[]{}

\author[orcid=0000-0002-1125-9187,sname='Hiramatsu']{Daichi Hiramatsu}
\affiliation{Center for Astrophysics , Harvard \& Smithsonian, 60 Garden Street, Cambridge, MA 02138-1516, USA}
\email[]{}

\author[orcid=0000-0002-3653-5598,sname='Gal-Yam']{Avishay Gal-Yam}
\affiliation{Department of Particle Physics \& Astrophysics, Weizmann Institute of Science, Rehovot 76100, Israel}
\email[]{}

\author[orcid=0000-0002-9392-9681,sname='Berger']{Edo Berger}
\affiliation{Center for Astrophysics , Harvard \& Smithsonian, 60 Garden Street, Cambridge, MA 02138-1516, USA}
\email[]{}

\author[orcid=0000-0002-4924-444X , sname='Bostroem']{K. Azalee Bostroem}
\affiliation{Steward Observatory, University of Arizona, 933 North Cherry Avenue, Tucson, AZ 85721-0065, USA}
\email[]{}

\author[orcid=0000-0002-1066-6098,sname='Chen ']{Ting-Wan Chen}
\affiliation{Graduate Institute of Astronomy, National Central University, 300 Jhongda Road, 32001 Jhongli, Taiwan}
\email[]{}

\author[orcid=0000-0003-4914-5625,sname='Farah']{Joseph Farah}
\affiliation{Las Cumbres Observatory, 6740 Cortona Drive, Suite 102, Goleta, CA 93117-5575, USA}
\affiliation{Department of Physics, University of California, Santa Barbara, CA 93106-9530, USA}
\email[]{}

\author[orcid=0000-0001-6395-6702,sname='Gomez']{Sebastian Gomez}
\affiliation{Center for Astrophysics , Harvard \& Smithsonian, 60 Garden Street, Cambridge, MA 02138-1516, USA}
\email[]{}

\author[orcid=0000-0001-5807-7893,sname='McCully']{Curtis McCully}
\affiliation{Las Cumbres Observatory, 6740 Cortona Drive, Suite 102, Goleta, CA 93117-5575, USA}
\email[]{}

\author[orcid=0000-0003-3643-839X,sname='Rho']{Jeonghee Rho}
\affiliation{SETI Institute, 189 Bernardo Ave., Ste. 200, Mountain View, CA 94043, USA}
\email[]{}





\begin{abstract}

Hydrogen-rich superluminous supernovae without narrow emission lines (SLSNe-II) are rare transients whose powering mechanisms remain debated, particularly the role of circumstellar medium (CSM) interaction. We present photometric and spectroscopic observations of SN 2019cqc to investigate its power source and progenitor mass loss. We model the lightcurve using MOSFiT with the \texttt{csm} and \texttt{csmni} models to constrain the explosion and CSM parameters. SN 2019cqc peaked at $M_g = -20.21 \pm 0.07$ and emitted $\sim 1.7 \times 10^{50}\,\mathrm{erg}$ of energy in the form of radiation. 
Photometric and spectroscopic modeling indicates that CSM interaction dominates in both scenarios, with best-fit CSM and ejecta masses of $\sim 4.5~ M_\odot$ and $\sim 32~ M_\odot$, respectively. The inferred CSM properties imply an extreme eruptive mass loss at a rate of $\sim 0.2–-0.3 ~M_\odot,\mathrm{yr^{-1}}$ in the years preceding the core collapse, consistent with a variable progenitor of luminous blue variable progenitor. We observe a persistent blueshifted asymmetry in the H$\alpha$ emission line. At early times ($\lesssim +110$~d), this is attributed to electron scattering with bulk Velocity of ejecta, while the profile at late epochs ($\gtrsim +402$~d onward) suggests the subsequent formation of dust in the ejecta. Additionally, we identify a distinct, short-lived feature at $\sim 4600$\,\AA, likely a blend of ionized C\,\textsc{iii}/N\,\textsc{iii} lines powered by the interaction of the SN-shock with the extended atmosphere.

\end{abstract}
\keywords{\uat{Type II supernovae}{1731} \uat{Circumstellar matter}{241} \uat{Stellar mass loss}{1613} \uat{Dust formation}{2269} \uat{Galaxies}{573} }


\section{INTRODUCTION}
Over the past two decades, untargeted surveys have uncovered a rare subclass of exceptionally luminous stellar explosions, superluminous supernovae (SLSNe). SLSNe are typically defined by a peak absolute magnitude brighter than $\sim -21$ mag \citep{2012Sci...337..927G,2019NatAs...3..697I}, although some authors have used a slightly lower threshold (e.g. $-20$ mag; 
\citealt{2016ApJ...830...13P,2019ARA&A..57..305G}). Regardless of the precise cutoff, SLSNe reach luminosities well above what normal core collapse supernovae (CCSNe) can produce. They are divided into hydrogen-poor (SLSN-I) and hydrogen-rich (SLSN-II) subclasses, by analogy with ordinary SNe I/II. SLSNe-II themselves are extremely rare; early 
estimates place their rate at only of order $\sim0.1\%$ of all CCSNe \citep{10.1093/mnras/stt213, 2014ApJ...792..135T}. 

Spectroscopically, SLSNe-II are not defined by any single distinctive feature \citep{hu2025sn2021aaevhydrogenrichsuperluminous}, but most show clear signs of strong circumstellar interaction \citep{2019ARA&A..57..305G,2021A&G....62.5.34N}. In particular, majority of known SLSNe-II exhibit narrow emission lines, mainly of Hydrogen, and are classified as SLSNe-IIn \citep{2024MNRAS.530..405S,2019ARA&A..57..305G}. 
SN~2006gy \citep{2007ApJ...659L..13O, 2007ApJ...666.1116S} is a prototype of this class. 
These narrow lines arise from slow-moving ($\sim100-150$ km\,s$^{-1}$), photoionized CSM around the star \citep{1990MNRAS.244..269S, 10.1093/mnras/268.1.173}. When the fast SN ejecta collide with this dense H-rich CSM, they drive powerful shocks that efficiently convert kinetic energy into radiation, yielding the extraordinarily high luminosities of SLSNe-IIn \citep{2008ASPC..401..185C, 1994ApJ...420..268C}. Early spectra of such events often show additional “flash-ionization” features (e.g.\ narrow He~II $\lambda4686$) from very recently ejected circumstellar medium (CSM), which appear within hours to days after explosion and fade by peak \citep{2014Natur.509..471G, Khazov_2016, Bruch_2023}. A much smaller subset of SLSNe-II shows broad Balmer lines from the ejecta in their early photospheric phase. Importantly, they do not exhibit the temporal evolution of the narrow emission line \citep{2018MNRAS.475.1046I, 2022MNRAS.516.1193K}. SN~2008es is widely considered as a prototype of these events \citep{Gezari_2009,2009ApJ...690.1303M}. 

For both SLSNe-II, and SLSNe-IIn, the CSM interaction has been widely considered the most plausible explanation to power the SN: the interaction luminosity can reach the required values via shock diffusion through an optically thick envelope \citep{1994ApJ...420..268C}. In practice, most well-studied SLSN-II lightcurves are consistent with models invoking massive CSM shells (often several $M_\odot$) around the progenitor \citep{2019ARA&A..57..305G, Pessi_2025}. Other models consider the presence of 
an additional central engine. For example, the spin-down of a newborn millisecond magnetar  
can inject energy into the ejecta to thermalize it  \citep{1971ApJ...164L..95O, 2010ApJ...717..245K, 2010ApJ...719L.204W}. SNe 2008es, 2013hx, PS15br \citep{2018MNRAS.475.1046I}, which did not show any narrow emission from the unshocked CSM, were explained as powered by a spin-down magnetar. 
However, all of these also showed signs of interaction $-$ The late-time spectra of SNe 2013hx $\&$  2015br show asymmetric, multi-component H$\alpha$ profiles, indicative of a non-spherical CSM, while SN2008es showed a UV excess \citep{2009ApJ...690.1303M}.

All scenarios for SLSNe generally invoke extremely massive progenitors. Models often require the mass of the progenitor stars of the order $\gtrsim30-100$ M$_\odot$ 
or more to achieve the observed energy and mass budget \citep{ 2017ApJ...835...13J,2018NatAs...2..887L}. In the cases of SLSNe-II, the progenitor stars retain a substantial amount of hydrogen in their envelopes until their collapse and are surrounded by dense CSM produced due to eruptive phenomena before the final explosions.
This implies to very high pre-SN mass-loss rates. Observational studies of Type~IIn SNe infer mass-loss rate of the order of $10^{-2}$ $-$ $1\,M_\odot$yr$^{-1}$ during the last few years before core collapse \citep{Smith_2014, Smith2017}, far above the $\lesssim10^{-4}\,M_\odot$yr$^{-1}$ expected from steady radiatively-driven winds \citep{2006ApJ...645L..45S, Smith_2014}. Such extreme mass-loss suggests that the eruptive mechanisms are similar to the giant eruptions observed in luminous blue variable (LBV), rather than a steady wind. 
This mechanism typically requires a CSM having a mass of the order of $\gtrsim 5 -$ 10 M$_\odot$  surrounding the star \citep{2007Natur.450..390W,2007ApJ...671L..17S,  2020NatAs...4..893N}. 

 When the CSM is optically thick, the primary luminosity peak may instead be powered by diffusion of shock-deposited energy (a delayed shock breakout), as invoked for SN~2006gy (\citealt{2007ApJ...671L..17S, Chevalier_2011} and references therein).
Whereas the variety in CSM structures leads to a wide range of observable effects after the explosion, particularly bumps in the lightcurves. 
However, compared to SLSNe-II (with and without narrow lines), the occurrence of bumps in the lightcurves of SLSN-I has been investigated in greater detail \citep{10.1093/mnrasl/slv210, 2019ARA&A..57..305G}. 
 Post-peak variability has also been identified in a number of SLSN-I events \citep[and references therein]{2016A&A...596A..67R, 2017ApJ...848....6Y, 2022ApJ...933...14H,2023ApJ...943...41C}, while recent work \citep{Pessi_2025} reports similar undulations in SLSN-II events, although their physical origin is not yet well constrained.  
 In general, the geometry and density profile of the CSM may produce asymmetric line profiles and peculiar shapes (e.g., bumpiness) of the 
 lightcurves $-$ well observed in iPTF13Z, a type IIn event \citep{2017A&A...605A...6N}. A similar scenario is also expected as the cause for the undulation in the lightcurves of SLSNe-II. 
 Alternatively, intermittent flares from a central engine \citep{10.1093/mnras/stx1028} could imprint variability on the lightcurve. In fact, recent statistical analysis also suggests that there is no obvious reason to consider SLSNe-II and Type IIn SNe to be produced through two completely different progenitor channels \citep{2024arXiv241107287H}.   

An important byproduct of CSM interaction is the formation of dust. Many interacting SNe (including SLSNe--I \& II) showed evidence of new dust in their post-shock gas, seen as infrared excesses and progressive blueshifts of emission lines \citep{2008ApJ...686..467S, 2021arXiv210907942C}; contributing a significant amount of dust to their environments \citep{ 2014Natur.511..326G, 10.1093/mnras/stz679}. SN~2020wnt is another SLSN--I that showed dominant coolant CO emission, and rising K-band continuum, indicating formation of hot and new dust in its ejecta \citep{tinyanont23}. Similar dust formation signatures have also been reported in the SLSN--I SN~2017egm \citep{Zhu_2023}.
 
Blueshifts of emission peaks may also arise from an asymmetric explosion with emission is powered by CSM interaction, in which electron scattering within the CSM produces the observed blueshift 
(\citealt{2014ApJ...797..118F} and references therein). 

In this work, we present the evolution of the SN~2019cqc and place it in the framework of SLSN-II \& SLSN-IIn as discussed above. The SN was initially analyzed by \citet{2022MNRAS.516.1193K} as SLSN-II. In this work, we revisit the event with a larger data set to explore its characteristics and evolution till the late epoch. The observation and data reduction are described in \S\ref{Discovery and Follow-up}, the extinction and distance estimates are detailed in \S\ref{extinction and distance cal}. The photometric and spectroscopic evolutions are presented in 
\S\ref{Photometric evolution of SN2019cqc} \& \S\ref{Spectroscopic evolution of SN2019cqc}, respectively. A detailed discussion on the lightcurve and spectral evolution is presented in \S\ref{Discussion}, and finally we have concluded in \S\ref{conclusion}. 

\begin{figure}
    \centering
    \includegraphics[width=1\linewidth]{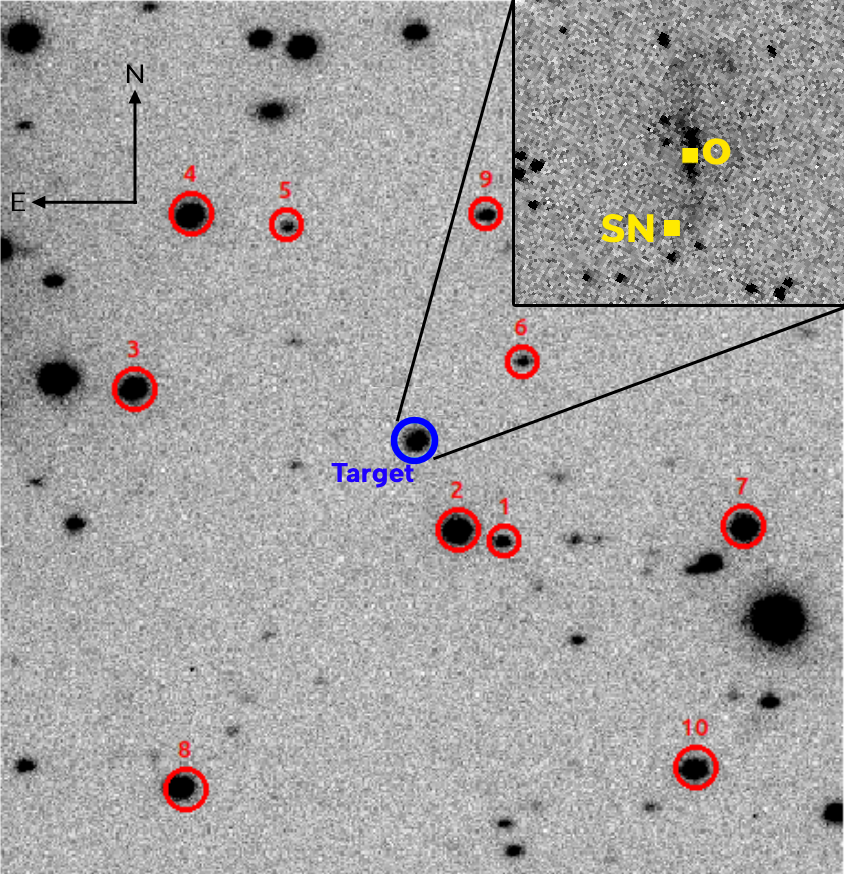}
    \caption{ $r$-band image of the field of SN~2019cqc obtained with the 1-m LCO telescope, covering $\sim3' \times 3'$ . The target (supernova + host) is marked with a blue circle and the reference stars with red circles. The inset shows an {\it HST} image of the host galaxy taken at MJD 60296.378, with the galaxy nucleus (O) and supernova position (SN) indicated. }
    \label{fig:Field of SN2019cqc}
\end{figure}

\section{Discovery and Follow-up Observations of SN~2019cqc}
\label{Discovery and Follow-up}
SN~2019cqc\footnote{\url{https://www.wis-tns.org/object/2019cqc}} (ZTF19aamrais) was first discovered by the Zwicky Transient Facility (ZTF) survey \citep{2019TNSTR.477....1N} on March 25, 2019 (MJD 58567.507) at a magnitude of $ \textit{g} = 19.28 \pm 0.13$ in an uncatalogued host galaxy at coordinates RA = 18:21:43.052 and Dec = +30:59:33.49. The transient was independently
discovered by ATLAS\footnote{\url{https://atlas.fallingstar.com/}} (ATLAS19ice; MJD 58603.562), Gaia\footnote{\url{http://gsaweb.ast.cam.ac.uk/alerts/home}} (Gaia19bsc; MJD 58573.598), and Pan-STARRS\footnote{\url{https://www2.ifa.hawaii.edu/research/Pan-STARRS.shtml}} (PS19ety; MJD 58719.365) sky surveys. The supernova was spectroscopically classified as a Type~II SLSN, with a redshift of $z=0.117\pm0.0003$, by the ZTF team \citep{2019TNSCR2814....1F}. Compared to \citet{2022MNRAS.516.1193K}, our observations provide broader wavelength coverage, including additional \textit{B}, \textit{V}, ATLAS  \textit{c} \&  \textit{o} , and UV bands, along with extended late-phase spectroscopy up to +478\,d after the \textit{r}-band peak. The photometric and spectroscopic follow-up observations are described in the following subsections.

\subsection{LCO Observations in \textit{B}, \textit{V} , \textit{g}$'$, \textit{r}$'$, and \textit{i}$'$ Bands}
We obtained 21 epochs of photometric observations of SN~2019cqc with the 1-m telescope, which is located at McDonald Observatory\footnote{\url{https://mcdonaldobservatory.org/}}, and a part of the Las Cumbres Observatory (LCO) network\footnote{\url{https://lco.global/}}. The LCO data were obtained through the Global Supernova Project (GSP) collaboration. Observations spanned from MJD 58617.441 ($\sim$10 days post-peak) to MJD 58807.051 ($\sim$200 days post-peak)in the \textit{g}$'$, \textit{r}$'$, and \textit{i}$'$ bands. For the \textit{r}$'$ and \textit{i}$'$ bands, we combined two 200-second exposures per epoch; for the \textit{B,V} and \textit{g}$'$ band, we combined two 300-second exposures. The field of SN~2019cqc is shown in Figure~\ref{fig:Field of SN2019cqc}. The SN is unresolved from its host galaxy. 
  Bias-subtraction, flat-fielding, cosmic-ray removal (using Laplacian kernel detection method;  \citealt{2001PASP..113.1420V}), alignment, and determination of mean FWHM and ellipticity in all object frames were carried out using the standard tasks of the data reduction software \texttt{IRAF}\footnote{\texttt{IRAF} stands for Image Reduction and Analysis Facility (IRAF), distributed by the National Optical Astronomy Observatory, which is operated by the Association of Universities for Research in Astronomy, Inc., under a cooperative agreement with the National Science Foundation.}. 
Image subtraction 
was also carried out using an \texttt{IRAF} and 
\texttt{DAOPHOT-II} \citep{1987PASP...99..191S} based code developed by  \citet{2011MNRAS.414..167R} (also described in \citealt{2012PhDT.......317R}). Finally, 
we performed PSF photometry using tasks in \texttt{DAOPHOT-II} to derive instrumental magnitudes.

We calibrated these magnitudes to the standard system using ten field stars (Figure~\ref{fig:Field of SN2019cqc}, Table~\ref{tab:refstar} ). As no SDSS standard stars were available, we adopted Pan-STARRS1 magnitudes of the reference stars, converted them to the SDSS system and Johnson \textit{BV} systems following \citet{2012ApJ...750...99T}, and used these for calibration. The calibrated magnitudes of SN~2019cqc obtained from LCO observations at different epochs are tabulated in Table~\ref{tab:lco_photometry} .
The inset of Figure~\ref{fig:Field of SN2019cqc} shows the HST image of the host galaxy of SN~2019cqc, where the host appears to be a spiral galaxy. This image was taken on MJD 60296.378 in the F336W filter, corresponding to $\sim1689$ d post-peak in the observer frame. We measure an angular separation of $1.24^{\prime\prime}$ between the transient and the host-galaxy nucleus, corresponding to a projected separation of $\sim3.3\,\mathrm{kpc}$, consistent with the spatial offsets observed for SLSNe-II \citep{Pessi_2025}.

\subsection{LT Observations in \textit{g}, \textit{r}, \textit{i}, and \textit{z} Bands}
SN~2019cqc was also observed with the Liverpool Telescope (LT) using the IO:O instrument\footnote{\url{https://telescope.livjm.ac.uk/TelInst/Inst/IOO/}} in the SDSS \textit{g}, \textit{r}, \textit{i}, and \textit{z} bands. The data comprise 19 epochs from MJD 58591.100 to 58714.080 (covering up to 107 days post-peak in the observer frame), with one 60-second exposure per epoch per filter. The LT data were reduced and calibrated using the same procedures as for the LCO observations. The calibrated magnitudes of LT follow-ups at different epochs are tabulated in Table~\ref{tab:lt_photometry}

\subsection{\textit{Swift} UVOT Observations}
SN2019cqc was observed with the Ultra-violet/Optical Telescope (UVOT) onboard the Neil Gehrels \textit{Swift} Observatory in the  \textit{uvw2}, \textit{uvm2}, \textit{uvw1}, \textit{u}, \textit{b}, and \textit{v}  bands, spanning MJD 58625.321 to 58654.417. The UV and \textit{u} bands were observed over four epochs. Both the  \textit{b} and \textit{v} bands were observed in a single epoch. All photometry was reduced using the standard pipeline within the \texttt{HEASOFT} software package. For 
source extraction, a small aperture of radius 3.5 arcsec was used to minimize the host contamination, while an aperture of radius 50 arcsec was used to determine the background\footnote{The UVOT data reduction was performed following the procedure detailed in \citet{2024MNRAS.528.6176R}.}. The calibrated magnitudes of the transient in {\it Swift} bands at different epochs are are tabulated in Table~\ref{tab:swift_19cqc} .The field was also observed few more times until MJD 59804.098 in {\it Swift}-bands; however, the transient was not detected in any of those observations and therefore not used in this work.

\subsection{ATLAS Observations in $\textit{c}$ and $o$ Bands}
SN~2019cqc was also monitored by the Asteroid Terrestrial-impact Last Alert System \citep[ATLAS;][]{2018PASP..130f4505T}. ATLAS uses two broad filters: the cyan (\textit{c}) band (4200--6500~\AA), approximately spanning the spectral-window equivalent to the  SDSS \textit{g+r} bands, and the orange (\textit{o}) band (5600--8200~\AA), similar to SDSS \textit{r+i}. The \textit{c}-band observations were obtained at 18 epochs spanning between MJD~58575.633 to 58803.213, while in \textit{o}-band the observations were performed between MJD~58565.450 and 58817.195 (nearly 55 epochs).

\subsection{ZTF \textit{g}- and \textit{r}-band Observations}
We supplemented our dataset with \textit{g}- and \textit{r}-band photometry from the ZTF Lasair database\footnote{https://lasair-ztf.lsst.ac.uk/}, covering MJD 58567.507 to 58717.311. These archival data significantly improve the temporal coverage of the lightcurve. Crucially, the ZTF data cover the early rising phase, which is sparsely sampled by the LT and LCO datasets. 

\subsection{Spectroscopic Observations}
Spectroscopic observations of SN~2019cqc were performed using three facilities: the 200-inch Hale Telescope  (P200) equipped with the Double Spectrograph (DBSP), the ESO New Technology Telescope (NTT) with the EFOSC2 instrument\footnote{  The NTT/EFOSC2 data were obtained under ePESSTO+ collaboration, an extended version of the spectroscopic survey ``Public European Southern Observatory Spectroscopic Survey of Transient Objects'' (PESSTO) consortium \citep{2015A&A...579A..40S}.}, and the Multiple Mirror Telescope (MMT). All the data were reduced using standard \texttt{IRAF} based pipelines. We obtained comparatively higher-resolution spectra at five epochs; two publicly available  P200/DBSP spectra were retrieved from the WISeREP \footnote{\url{https://www.wiserep.org}} archive \citep{2012PASP..124..668Y} (also analyzed by \citealt{2022MNRAS.516.1193K}), and three additional spectra were obtained with the MMT.
The remaining medium-resolution spectra were obtained with the ESO-NTT. For the epochs with multiple exposures, first, we combined the spectra to improve the signal-to-noise ratio and then proceeded to carry out further analysis.  The log of spectroscopic observations is presented in Table~\ref{tab:obslog} .

\section{DISTANCE AND EXTINCTION TOWARD SN~2019cqc}
\label{extinction and distance cal}
An accurate assessment of the extinction and distance toward the SN is necessary to generate its bolometric lightcurve and to probe different physical characteristics of the event. Redshift ($z$) of the SN, measured from the shift in the spectral line, is $0.1173 \pm0.0003$. Under a $\Lambda$CDM cosmology, this corresponds 
to a luminosity distance $D_L = 550\pm1$ Mpc, considering $\Omega_M = 0.286$, $\Omega_{\text{vac}} = 0.714$,  and $H_0 = 69.6$ $\mathrm{km\,s^{-1}\,Mpc^{-1}}$ 
\footnote{\url{https://www.astro.ucla.edu/~wright/CosmoCalc.html}} \citep{2014ApJ...794..135B}.
This implies that the distance modulus ($\mu$) of the SN is $\sim$ 38.7. 
An approximate K-correction ($= -2.5 \log(1+z)$) was applied when computing the absolute magnitude in each band \citep{hogg2002kcorrection,2023ApJ...943...41C}.

The galactic reddening E($B-V$) along the line-of-sight of SN~2019cqc is 0.103$\pm $0.004 mag, as determined by the 100 $ \mu m$ all-sky dust extinction map \citep{Schlafly_2011}. 
The extinction due to SN host galaxies is often estimated from the SN spectra using the equivalent width ($EW$) of Na\,\texttt{I}~D
 absorption feature \citep{Turatto2003}. However, we were unable to clearly discern the Na\,\texttt{I}~D impression due to host in any of the available spectra of SN~2019cqc. 

The most noticeable characteristic, found in the +110 day spectrum around the wavelength range where the host's Na\,\texttt{I}~D is expected, has $EW \leq 1.29~\text{\AA}$. 
 Whereas, the Na\,\texttt{I}~D absorption feature due to the Milky Way (MW), found in the same spectrum, has an EW roughly 1.68$\pm $0.7 \AA\,, where the uncertainty in the EW has been estimated using  equation~(6) of \citet{2006AN....327..862V}. Adopting the empirical relation proposed by \citet{Turatto2003}, we obtain the corresponding value of reddening of the MW as $E(B-V) = 0.26 \pm 0.11$. This result is consistent with the galactic reddening derived from \citet{Schlafly_2011}.

Thus for this work, we have assumed that total reddening along the line-of-sight is mainly due to the MW, and the host contribution is negligibly small. This corresponds to a total extinction towards SN~2019cqc in $V$-band is $A_V \approx 0.312$ mag, adopting a uniform value of the total-to-selective extinction ratio (R$_V = 3.1$) in the $V$-band for the MW.

\begin{figure*}
    \centering
    \includegraphics[width=1\textwidth]{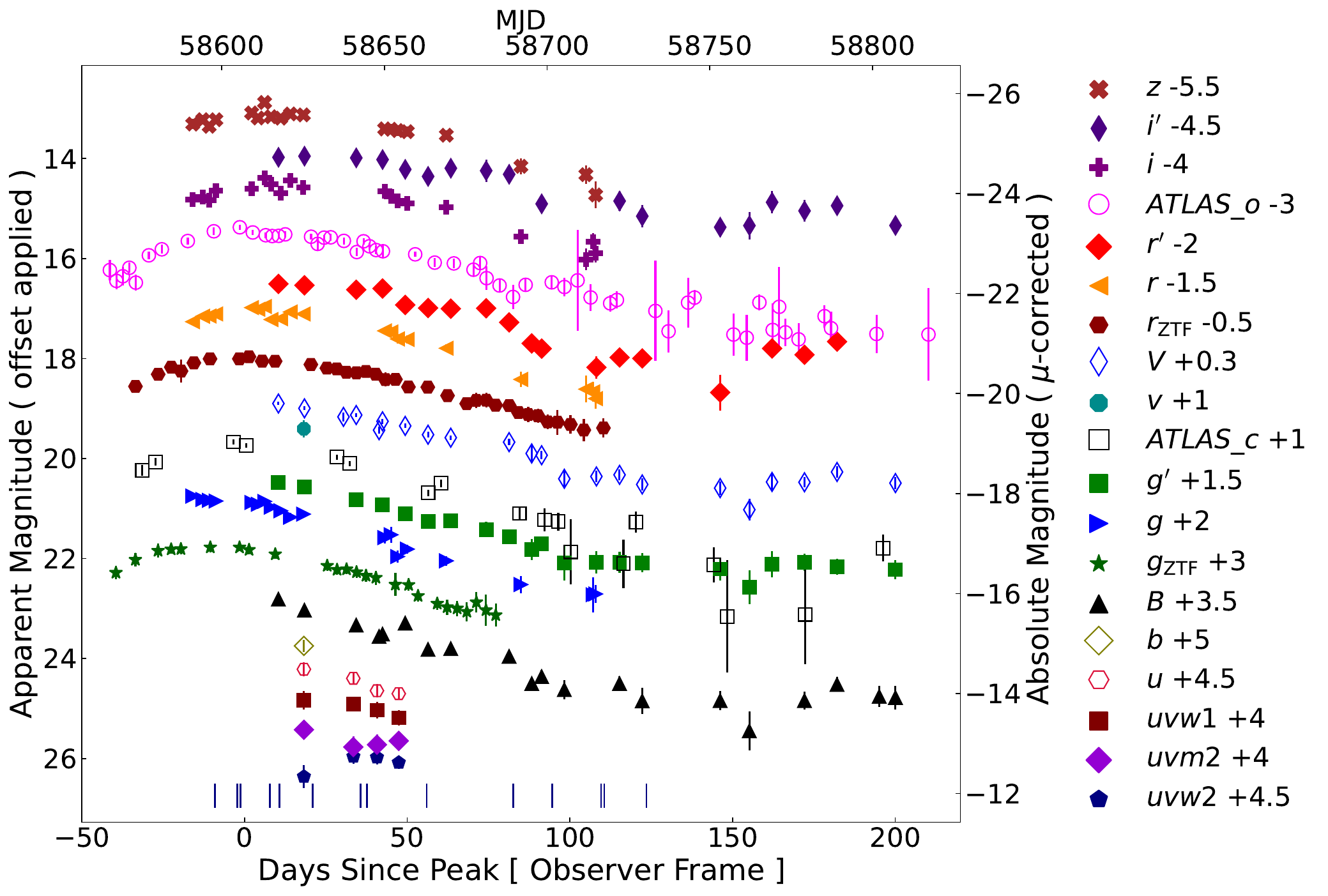}
    \caption{Multiband apparent lightcurves of SN~2019cqc obtained from ground-based and \textit{Swift} observations. The secondary axis on the right shows the absolute magnitudes, corrected only for  distance modulus ($\mu$ = 38.7 mag)}. Vertical ticks mark the epochs of spectroscopic observations.
    \label{fig:apparentmagnitude}
\end{figure*}

\section{Photometric evolution of SN~2019cqc}
\label{Photometric evolution of SN2019cqc}

The multi-band lightcurves of SN~2019cqc are shown in Figure \ref{fig:apparentmagnitude}. The lightcurves peak early in lower wavebands (e.g., ZTF \textit{g}-band) than in  longer wavebands (e.g., ZTF \textit{r}-band). 
The ZTF \textit{r}-band data, which cover the rise to maximum, were fitted with a polynomial function, giving roughly MJD 58607.3 as the epoch of peak brightness. This peak-brightness epoch in the ZTF \textit{r}-band has been adopted as the reference point of the time axis (t$_0$) throughout the analysis. 
Noteworthy, while studying the photometric properties of SN~2019cqc, \citet{2022MNRAS.516.1193K} considered the peak of the ZTF \textit{g}-band lightcurve as the origin of time (i.e., MJD 58595.2) in their analysis. Essentially, this implies that there is an offset of $\sim10$ days (in the rest frame) between the phases mentioned in this work and the corresponding phases adopted by them. Throughout this work, this offset has been considered while comparing our results with theirs.

The smooth rise in all optical bands is noticeable. However, during decay, they exhibit undulation at relatively late phases ($>+130$d after the ZTF \textit{r}-band maximum).  We note that at these late phases, 
evolution of SN~2019cqc was not densely followed by ZTF survey; while a possible rebrightening was observed in LCO \& ATLAS bands at around 168 days after maximum. 
  
Similar fluctuations have also been observed in several supernovae at comparable phases, where the ejecta-CSM interaction is thought to be the principal energy source.

\subsection{Evolution of Bolometric Luminosity}
\label{Evolution of Bolometric Luminosity}
\begin{figure}
    \centering
    \includegraphics[width=1\linewidth]{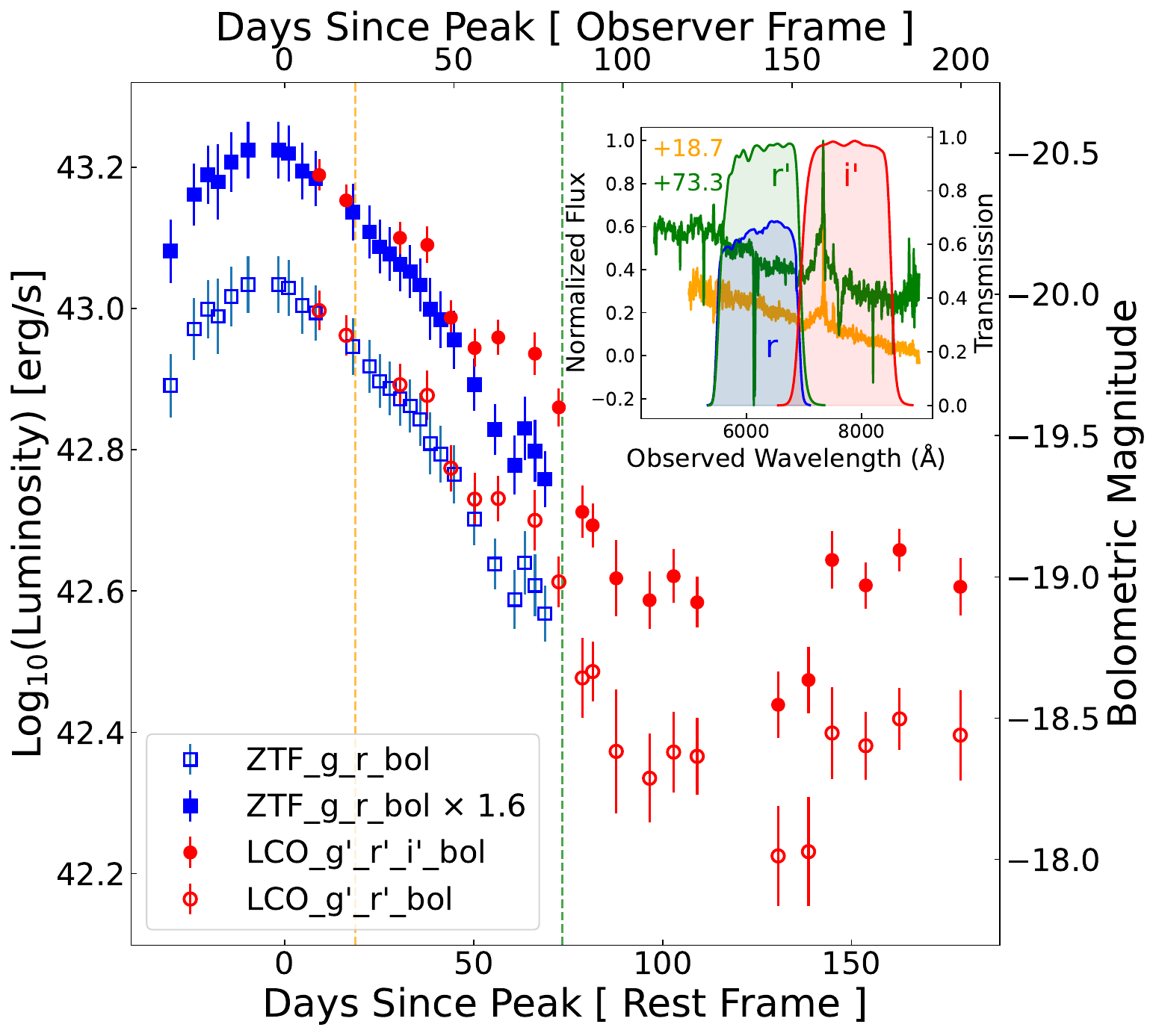}
    \caption{Pseudo-bolometric lightcurves of SN~2019cqc derived from ZTF $gr$ photometry (blue open symbols) and from \textit{g}$'$\textit{r}$'$\textit{i}$'$ observations (red). The $gr$ curve has been scaled by a factor of 1.6 to match the \textit{g}$'$\textit{r}$'$\textit{i}$'$ curve.  At later epochs, the $gr$ luminosity underestimates the total flux due to the increasing contribution of the \textit{i}$'$ band. The inset shows the \textit{r}$'$, $r$, and \textit{i}$'$ filter transmission curves over spectra obtained at +18 and +73 days, illustrating how the broadening of H$\alpha$ enhances the flux within the \textit{i}$'$ band.  A vertical offset has been applied to the +18d spectrum to show the H$\alpha$ features clearly.}  \label{fig:gr_gri_bol_lum_resp_curve_spec.pdf}
\end{figure}

To construct the pseudo-bolometric lightcurve, we first corrected the observed magnitudes for extinction and integrated the flux in the \textit{g}$'$, \textit{r}$'$, and \textit{i}$'$ bands. To create a complete pseudo-bolometric lightcurve from the rising phase to the late decline phase, we compared ZTF \textit{g},\textit{r}-band measurements with LCO \textit{g}$'$,$\textit{r}~'$,$\textit{i}~'$-band measurements. In epochs where the two datasets overlap, the pseudo-bolometric luminosities constructed from LCO \textit{g}$'$,\textit{r}$'$-bands show excellent agreement with those constructed from ZTF \textit{g},\textit{r}-bands, confirming a consistent calibration in both filter systems. We computed the ratio between pseudo-bolometric luminosities of the LCO \textit{g}$'$,\textit{r}$'$,\textit{i}$'$ to LCO \textit{g}$'$,\textit{r}$'$-bands during the overlap epochs and found that this ratio evolves with time. During the early epochs of the overlapping phases, when the spectrum is continuum-dominated, the median ratio is $\sim$1.6. We therefore scaled the ZTF \textit{g},\textit{r}-band pseudo-bolometric luminosities using this empirically determined ratio to estimate the pseudo-bolometric luminosities in optical bands (\textit{g},\textit{r},\textit{i}-band). This approach yields a smoothly connected pseudo-bolometric lightcurve that matches the LCO \textit{g}$'$,\textit{r}$'$,\textit{i}$'$ measurements very well during the overlap period (see Figure \ref{fig:gr_gri_bol_lum_resp_curve_spec.pdf} ). After $\sim$50 days, however, the scaled ZTF $\textit{gr}$ luminosity falls below the LCO \textit{g}$'$,\textit{r}$'$,\textit{i}$'$ luminosity. This divergence is primarily due to the increasing contribution of the \textit{i}$'$ band, which coincides with the emergence of a broad H$\alpha$ component. As the H$\alpha$ emission develops, particularly in its wings $-$ a larger fraction of the total flux is captured within the \textit{i}$'$ band. This is demonstrated in the inset of Figure \ref{fig:gr_gri_bol_lum_resp_curve_spec.pdf} ,
where the \textit{r}$'$, \textit{r}, and \textit{i}$'$ filter transmission curves are plotted over spectra obtained at +18 and +73 rest-frame days. While the +18 day spectrum shows only weak H$\alpha$ broadening, by +73 days the line has developed strong wings, significantly enhancing the flux in the \textit{i}$'$ band.

\section{Spectroscopic evolution of SN~2019cqc}
\label{Spectroscopic evolution of SN2019cqc}

\begin{figure*}
    \centering
    \includegraphics[width=\textwidth]{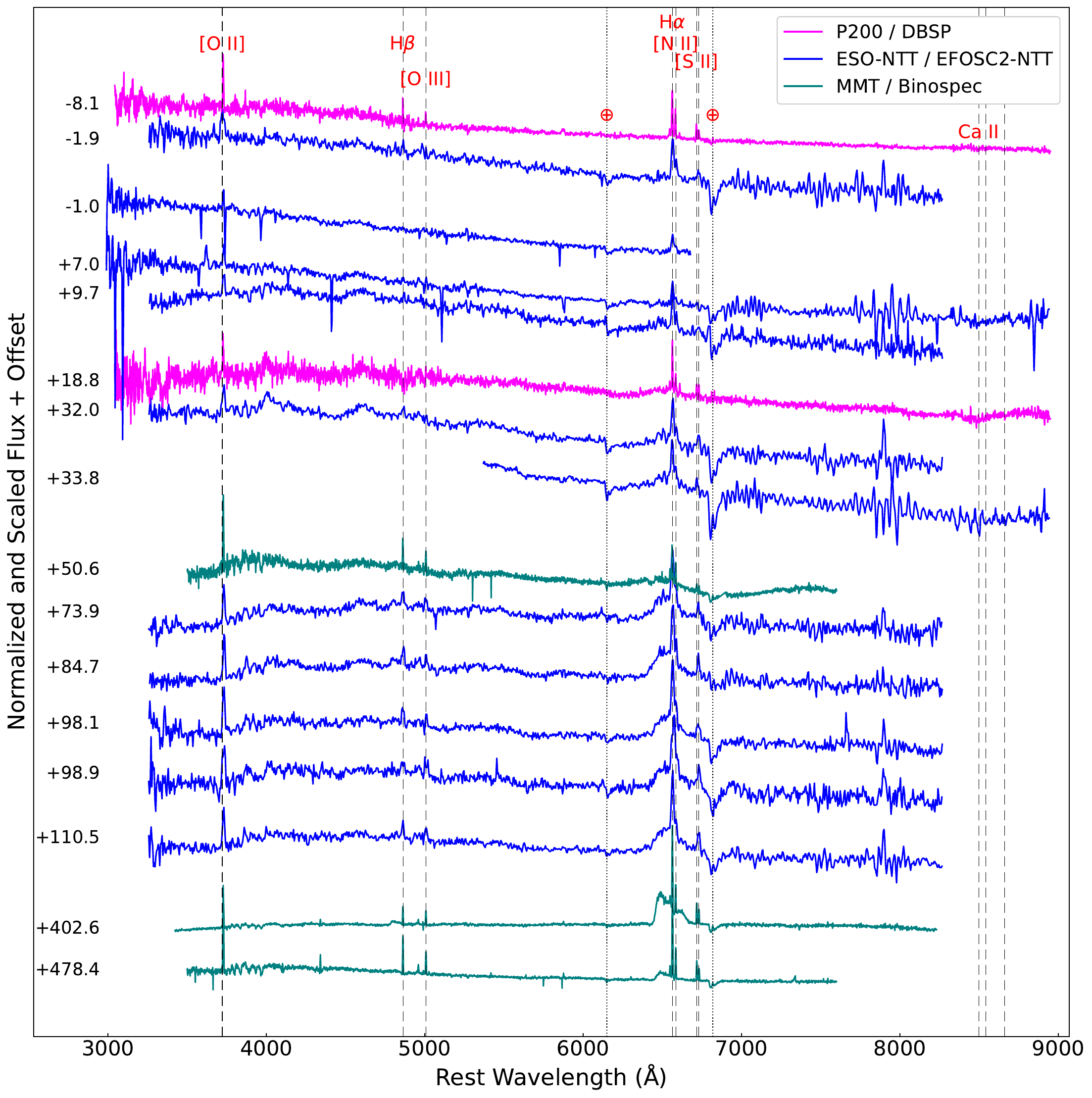}
    \caption{Spectral evolution of SN~2019cqc from $-8$d to $+478$d. The spectra are displayed in the source rest frame and corrected for extinction. Emission lines of the host galaxy are marked with vertical dashed lines, while the dotted lines mark the telluric absorption features.}
    \label{fig:spectral_ladder}
\end{figure*}

Figure~\ref{fig:spectral_ladder} presents the spectral sequence of SN~2019cqc over 15 epochs, spanning from $-8$ to $+478$ day relative to the peak \textit{r}-band brightness in the rest frame. The spectra have been corrected for redshift and Galactic extinction, and prominent features are labeled. Semi-regular fluctuations beyond $\sim8000$~\AA\ in some ESO-NTT spectra are attributed to fringing effects in the detector. Some of the major emission lines from the host galaxy (e.g., H$\alpha$, H$\beta$, [N\,\texttt{II}], [S\,\texttt{II}] \& [O\,\texttt{II}]) have been marked by the vertical dashed lines.  

Apart from the emission features at around $\sim$4000~\AA\ and 4600~\AA \ (see  \S\ref{feature_4600} for a further discussion), no significant broad emission lines have been observed in the early spectra of the transient, and the spectra  remained dominated by the narrow emission lines of the host galaxy. Only by +18d, broad lines start to appear prominently in H$\alpha$, which later became more asymmetric in shape. Additionally, the Ca~II triplet ($\lambda\lambda 8498.0, 8542.1, 8662.1$) is also observed at the red-wing of the +18d spectrum, as also reported by \citet{2022MNRAS.516.1193K}.

\section{Discussion} 
\label{Discussion}

\subsection{Possible fluctuation in the late-time lightcurves}

Bump-like features have been reported in the lightcurves of some SLSNe-II, although it is not very common. For e.g., while analyzing a sample of 107 hydrogen-rich SLSNe \citet{Pessi_2025} (visually) found lightcurve bumps only in 7 objects (nearly 6.5\%), out of which two (SNe 2018hsb, 2022pjl) were spectroscopically classified as SLSN-II, and five as SLSN-IIn. None of the SLSNe-II reported by \citet{2018MNRAS.475.1046I} \& \citet{2022MNRAS.516.1193K} showed post-maxima bumps in their lightcurves. Such undulations 
during the decline phase are attributed to the ejecta encountering inhomogeneities in the CSM,   
possibly generated by episodic mass loss from a LBV progenitor. On the other hand, \citet{2022ApJ...933...14H} found that within SLSNe--I, up to 76\% may 
exhibit light-curve undulations, which they suggest could either arise from fluctuations in the central engine (e.g., variability in magnetar energy input) or from shock-interaction with confined CSM structures such as shells or disks.

To place SN~2019cqc in this context, we compare its \textit{g}-band lightcurve with a representative set of SLSNe-II from the literature, along with few SLSN-IIn, for e.g., SN~2018bwr, which showed a broad bump and an exceptionally slow decline, requiring more than a year to fade to 10\% of its peak luminosity  (see left panel of Figure~\ref{fig:lccomp} ).Other noteworthy cases are two SLSNe-IIn (SNe 2021mz, 2022gzi), and one SLSN-II (SN~2022pjl), which exhibited secondary bumps after maximum light. 

This comparison suggests that there may be a possible late time bump in SN~2019cqc (at  $\sim$168 days after the peak in ZTF \textit{r}-band; see \S\ref{Bol_char}) that happened at a phase consistent with other SLSNe-II, and SLSNe-IIn, which showed bumps in their lightcurves. This implies that the late-time evolutions of these events are possibly similar.
In the \textit{g}-band, SN~2019cqc and SN~2020bfe evolved comparably, starting from the early phases, suggesting that their progenitor properties and powering mechanisms of these events are possibly similar.

This diverse sample essentially highlights that, by-and-large the structure of the CSM of SN~2019cqc is similar to most of the SLSNe-II, which exhibit bumps during the post-maximum decline phase, while only a small subset of events, such as SN~2018dfa and SN~2021lhy, have shown pre-maximum undulations.

\subsection{Characterization of Bolometric evolution}
\label{Bol_char}

\begin{figure}
    \centering
    \includegraphics[width=1\linewidth]{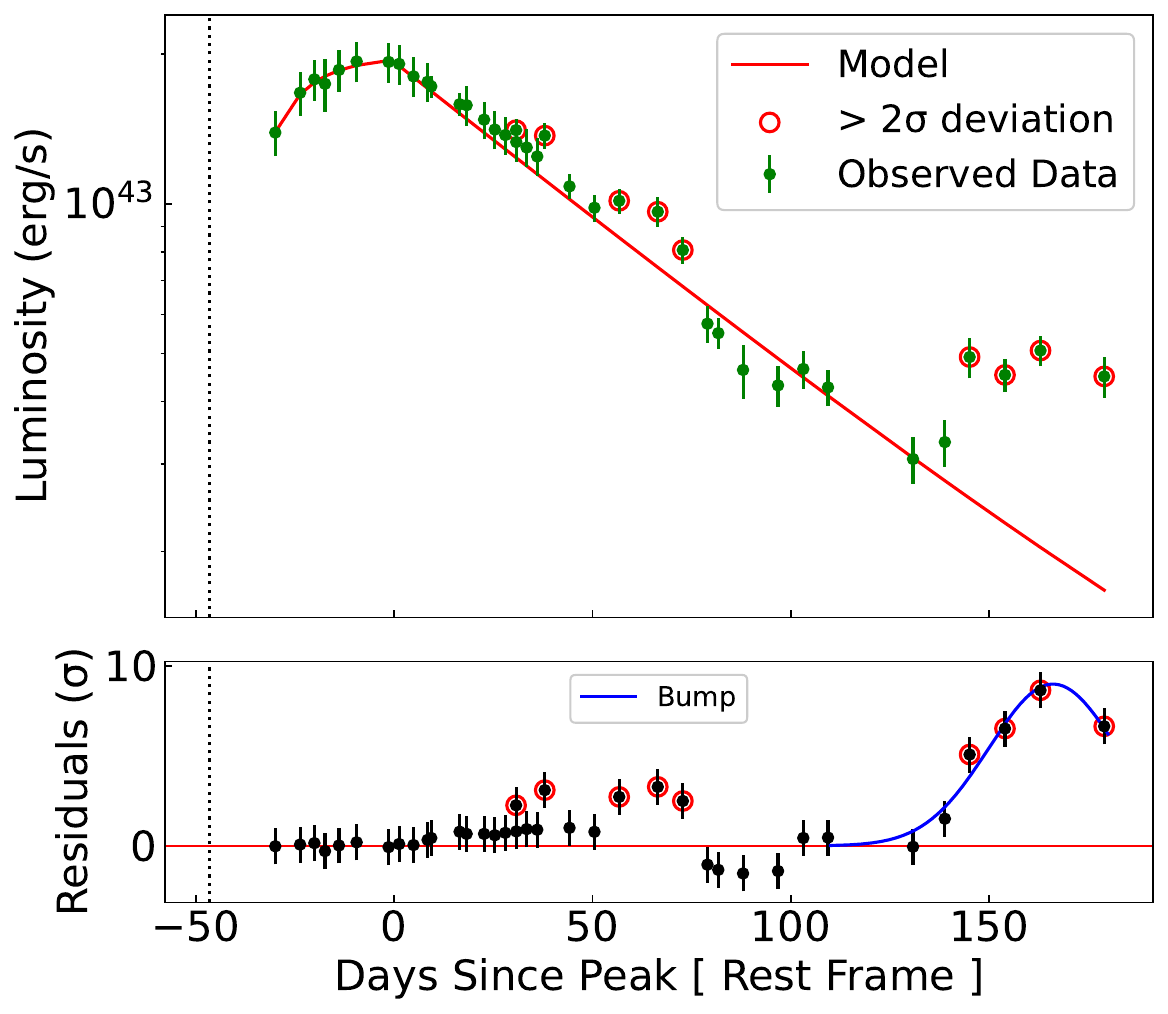}
    \caption{Pseudo-bolometric \textit{g}$'$\textit{r}$'$\textit{i}$'$ lightcurve of SN~2019cqc with analytic fits. 
\textit{Top:} observed luminosity (green points) and the best-fit model (solid line) shown on a logarithmic scale. Data points with $>2\sigma$ deviations from the model are highlighted with red circles and are interpreted as possible bumps. 
\textit{Bottom:} residuals in units of $\sigma$, defined as $(\log_{10}L_{\mathrm{data}} - \log_{10}L_{\mathrm{model}})/\sigma_{\log}$, with Gaussian profiles over-plotted on the probable rebrightening feature. 
Although the axis is logarithmic for clarity, all luminosities are expressed in linear units of erg~s$^{-1}$.}
    \label{fig:bol_lum_model_data.eps}
\end{figure}

\begin{figure*}
\centering

\subfigure{%
\includegraphics[width=0.47\linewidth]{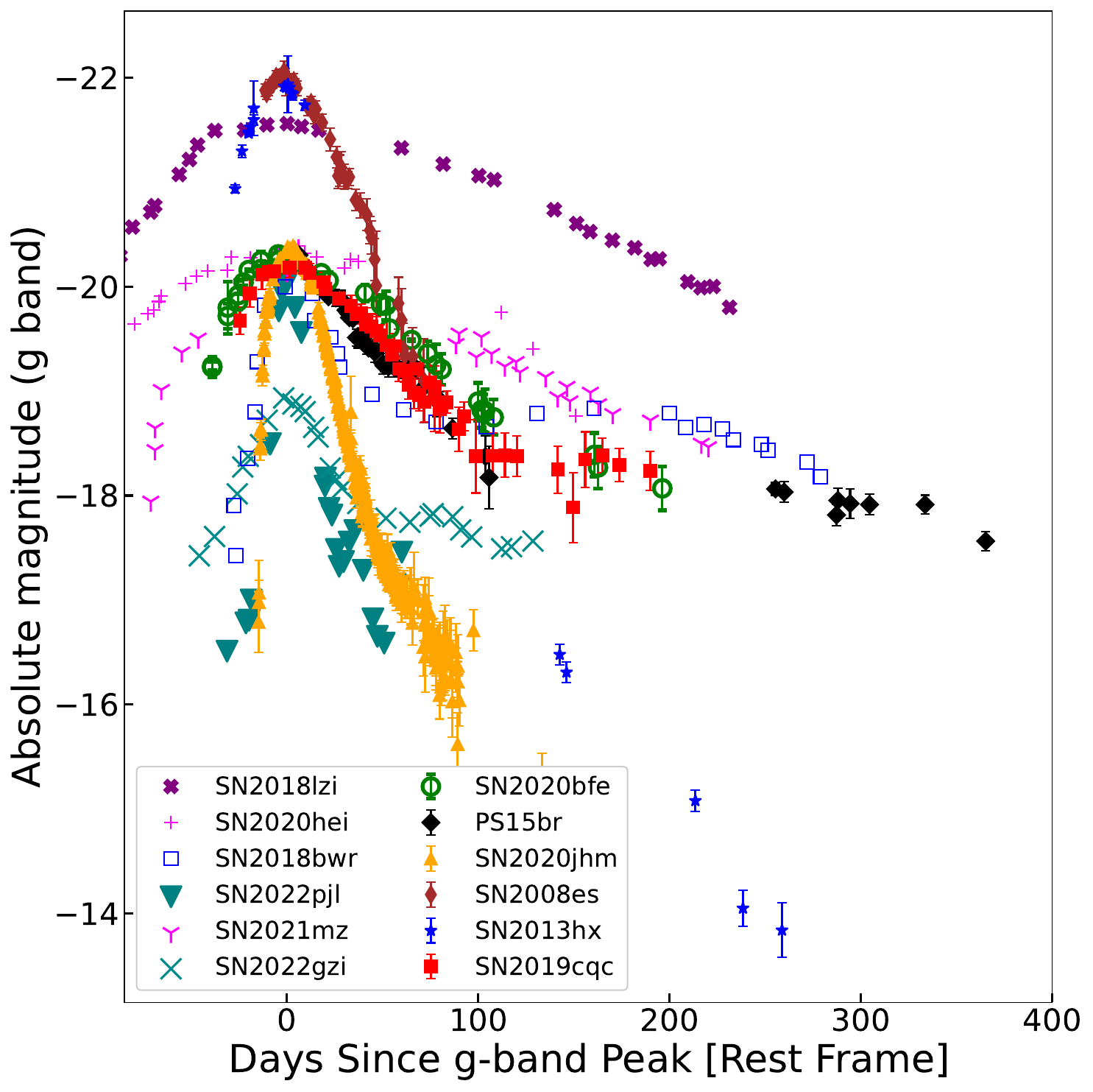}
\label{fig:abs}}
\hspace{0.0081\linewidth}
\subfigure{%
\includegraphics[width=0.47\linewidth]{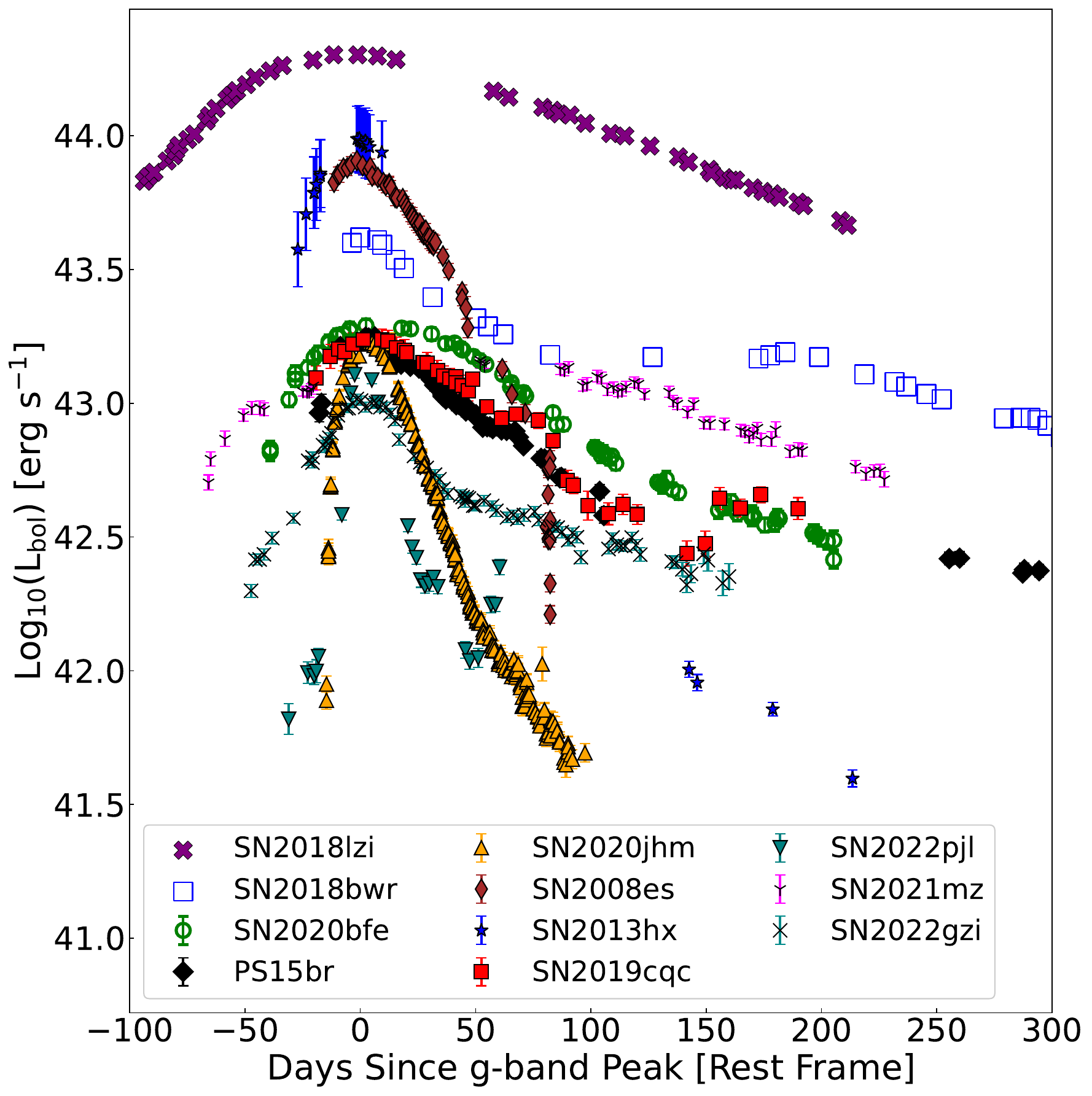}
\label{fig:bollum}}
\caption{
Left: \textit{g}-band absolute light curve of SN~2019cqc compared to a representative sample of SLSNe-II. 
Right: $\textit{gri}$ pseudo-bolometric light curve showing SN~2019cqc among the less luminous events.
    Data for comparison are taken from literature \citep{Gezari_2009,2018MNRAS.475.1046I,2022MNRAS.516.1193K,Pessi_2025}.
}
\label{fig:lccomp}
\end{figure*}

The rising phase of the \textit{g}$'$\textit{r}$'$\textit{i}$'$ pseudo-bolometric lightcurve of SN~2019cqc is modeled (see Figure \ref{fig:bol_lum_model_data.eps}) using an exponential function \citep{Ofek_2014} of the form
\[
L(t) = L_{\mathrm{max}}\left[1 - \exp\left(\frac{t_{\mathrm{exp}}- t}{t_e}\right)\right],
\]
where $L(t)$ is the luminosity at time $t$. $t_e$ is the characteristic timescale, and $t_{\mathrm{exp}}$ corresponds to the explosion epoch (the time when the luminosity approaches zero). All are measured in the rest frame of the SN. This prescription provides an estimate of the explosion epoch to be nearly $-46\pm4$ days relative to the \textit{r}-band peak. 

The decline phase of the lightcurve is modeled under the \texttt{TigerFit} framework\footnote{\url{https://github.com/manolis07gr/TigerFit}}, assuming a constant-density CSM \citep{2013ApJ...773...76C}.  Data points lying more than $2\sigma$ above the best-fit decline model are marked separately in the figure \ref{fig:bol_lum_model_data.eps}. These deviated points are not considered in the MOSFiT modeling of the lightcurves, as discussed in the subsequent section of this text. 
In addition to the overall smooth rise and decline, the pseudo-bolometric light curve shows two possible deviations from a smooth evolution. The first occurred on $\sim$60  day in rest frame.  
However, the sparse observational coverage during these epochs prevents any firm interpretation. \citet{2022MNRAS.516.1193K} also did report any bumpiness during this phase. 

The second, broader deviation peaks at $\sim$168~days with a FWHM of $\sim$43~days, and an amplitude of $7.5\sigma$. 
We note that at the comparable epoch \citet{2022MNRAS.516.1193K} did not find any bumpiness in the lightcurve of SN~2019cqc, where they have only one data point. On the other hand, our five consecutive observations showed an enhancement in luminosity from its regular decay trend. We assumed that this enhancement in flux, as observed by us, is real and marked this late-time undulation as a possible bump in the lightcurve of SN~2019cqc. However, one may consider this flux enhancement as a flattening in the lightcurve as well. Noteworthy, in both cases an extra power source is required to keep the SN more luminous than its constant decline trend. 
 
If this late-time broad bump is real, it is more plausibly linked to the interaction of SN ejecta with a clump, present in the extended CSM along the line-of-sight. 
This assumption essentially associates the encounter of forward shock with the clump at a rest-frame time of $t \simeq (168+46) \simeq 214$~days after explosion, which corresponds to a distance of $R \approx 1.85\times10^{16}$~cm, assuming an average shock velocity of $v_{\rm sh} \sim 10^{4}$~km~s$^{-1}$. Adopting $v_w \simeq 500$~km~s$^{-1}$, as the velocity of the erupted material, this would correspond to a mass-loss episode that occurred approximately $t_{\mathrm{pre}} \approx 11$~yr before the explosion, and may be associated with the formation of this clump.

Comparison of the pseudo-bolometric lightcurve of SN~2019cqc with that of other SLSNe-II \& SLSNe-IIn with few having post-maxima bumps (see the right panel of Figure~\ref{fig:lccomp} ),  demonstrates that SN~2019cqc is one of the less luminous members of the entire SLSN-II population. While SN~2018lzi is an exceptionally luminous event, radiating more than $10^{51}$~erg; most others fall within the range $\sim10^{51}$~erg. SN~2022pjl showed a sharp rise before maximum and a rapid decline after that.  
Unlike the $g$-band lightcurve, the bolometric luminosities of SN~2019cqc and SN~2020bfe did not evolve in the same way. Although the peak luminosities of these two SNe are comparable, SN~2019cqc declined slightly faster than SN~2020bfe after the maximum. Beyond +150d, a rise in the luminosity of SN~2019cqc is noticeable (like a bump in the lightcurve), whereas at a similar phase, SN~2020bfe (and other SLSNe-II without any bump) kept decaying with a constant rate. This indicates the presence of an extra power source, making SN~2019cqc more luminous during that phase.
We also note that the phase, when this probable bump in SN~2019cqc appeared, is consistent with phases of bumps in some of the SLSNe-II events like SNe 2021mz, 2022pjl. 
Overall, the sample shows that, depending on the CSM distribution \& density profile, SLSNe-II show a wide variety in the evolution of their bolometric luminosities. 

\begin{figure}{h}
    \centering
    \includegraphics[width=1\linewidth]{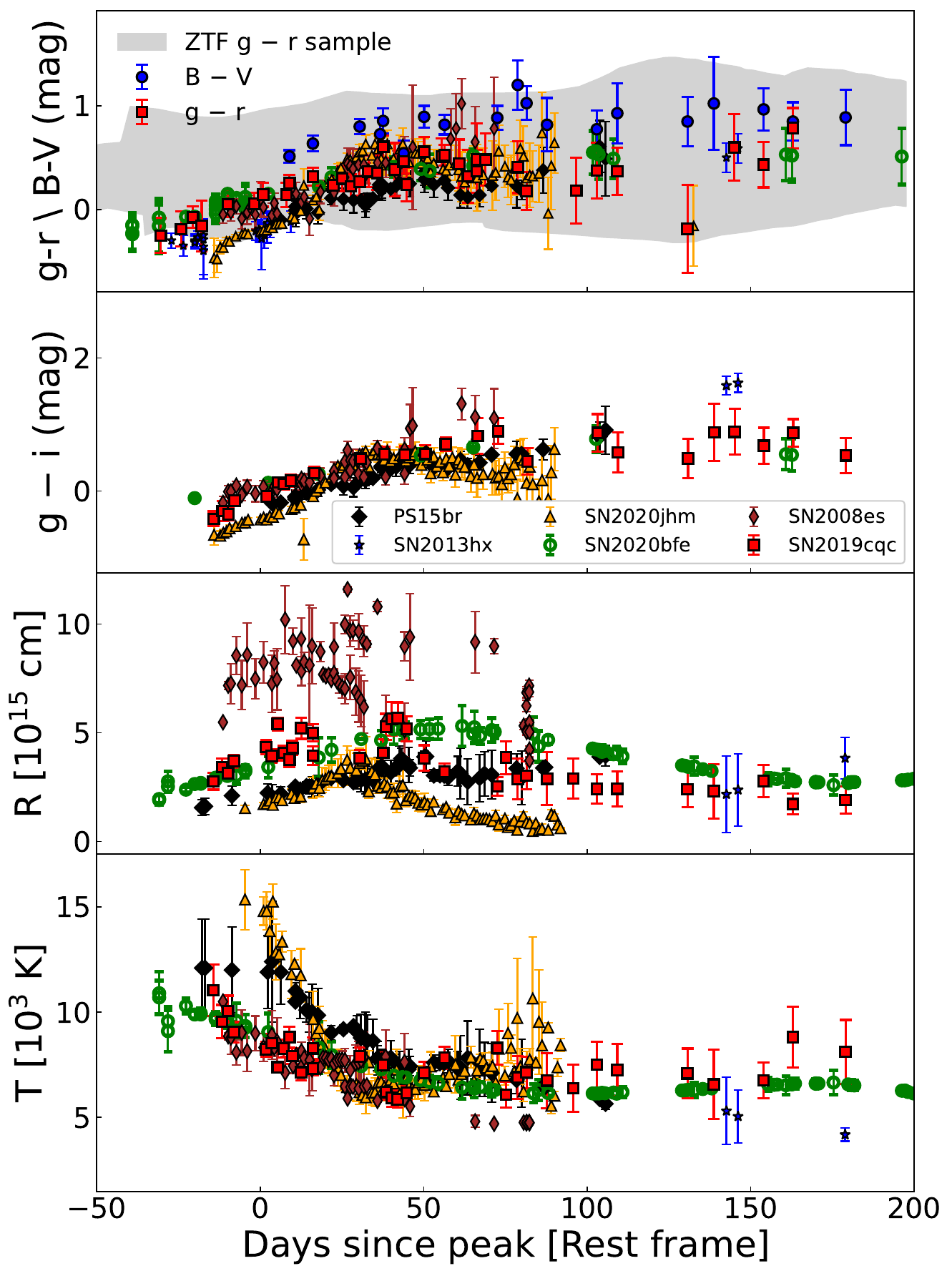}
    \caption{Comparison of the color (corrected for the adopted extinction), temperature, and radius of SN~2019cqc with the sample from \citet{Gezari_2009,2018MNRAS.475.1046I,2022MNRAS.516.1193K,Pessi_2025}. 
    }
    \label{fig:color}
\end{figure}

We observe that the $(\textit{g-r})$ color evolution is comparable to that of other SLSNe-II reported in the literature. \citep{Gezari_2009,2018MNRAS.475.1046I,2022MNRAS.516.1193K, Pessi_2025}. Overall, SN~2019cqc follows a redward trend similar to the comparison sample (see Figure \ref{fig:color}), consistent with the cooling of the ejecta. At the early epochs ($\lesssim$ 20 days before peak), the ($\textit{g-r}$) / ($\textit{B-V}$) color of our SLSN-II sample is little bluish (not very significant) in comparison to the ZTF sample (gray shaded region) adopted from \citet{Pessi_2025}.

The ($\textit{g-i}$), ($\textit{g-r}$), and ($\textit{B-V}$) colors exhibit a plateau-like evolution beyond $\sim$70 days.  Assuming a blackbody evolution of the SN, we estimate the temporal evolution of the photospheric radius ($R$) and temperature ($T$).  The inferred radius shows a plateau beyond $\sim$60 days (see Figure \ref{fig:color}), while the temperature remains approximately constant within the uncertainties during this phase. This departure from the monotonic cooling trend indicates the presence of an additional source of input energy that keeps the SN luminous at these epochs. Such behaviour could also be explained by interaction between the SN ejecta and previously ejected dense circumstellar material, which temporarily reheats the photosphere and delays its cooling. 

\subsection{Multiband lightcurve modelling with MOSFiT}
\label{Mosfit}
To estimate the physical parameters of the explosion, we carried out analytical multiband lightcurve modelling with the Modular Open-Source Fitter for Transients (MOSFiT\footnote{\url{https://github.com/guillochon/MOSFiT/tree/master}}; \citealt{Guillochon_2018})\footnote{MOSFiT is a Python-based framework that employs Monte Carlo ensembles of semi-analytical models to fit observed lightcurves and returns Bayesian posterior distributions for the model parameters}. 
As discussed in previous sections, overall photometric evolution and peak luminosity of SN~2019cqc are consistent with those of SLSNe-II events, which are supposed to be powered by interaction. The spectral features of the SN showed the signature of electron scattering and the asymmetric line profile in the late epoch (see \S\ref{Early Line Profile}, \S\ref{line profile decomposition}), consistent with the CSM interaction process. 
Therefore, we adopted MOSFiT models that include CSM interaction (along with radioactivity) as a power source.

In the MOSFiT \texttt{csm} model \citep{2012ApJ...746..121C,Villar_2017}, the luminosity results from the conversion of the kinetic energy at both the forward and reverse shocks into radiation. The \texttt{csmni} model augments this by including the heating from the radioactive decay of $^{56}\mathrm{Ni}$ in addition to CSM interaction. We modelled the lightcurve of SN\,2019cqc with both \texttt{csm} and \texttt{csmni} configurations. 

\begin{figure*}   
    \centering
    \includegraphics[width=\textwidth]{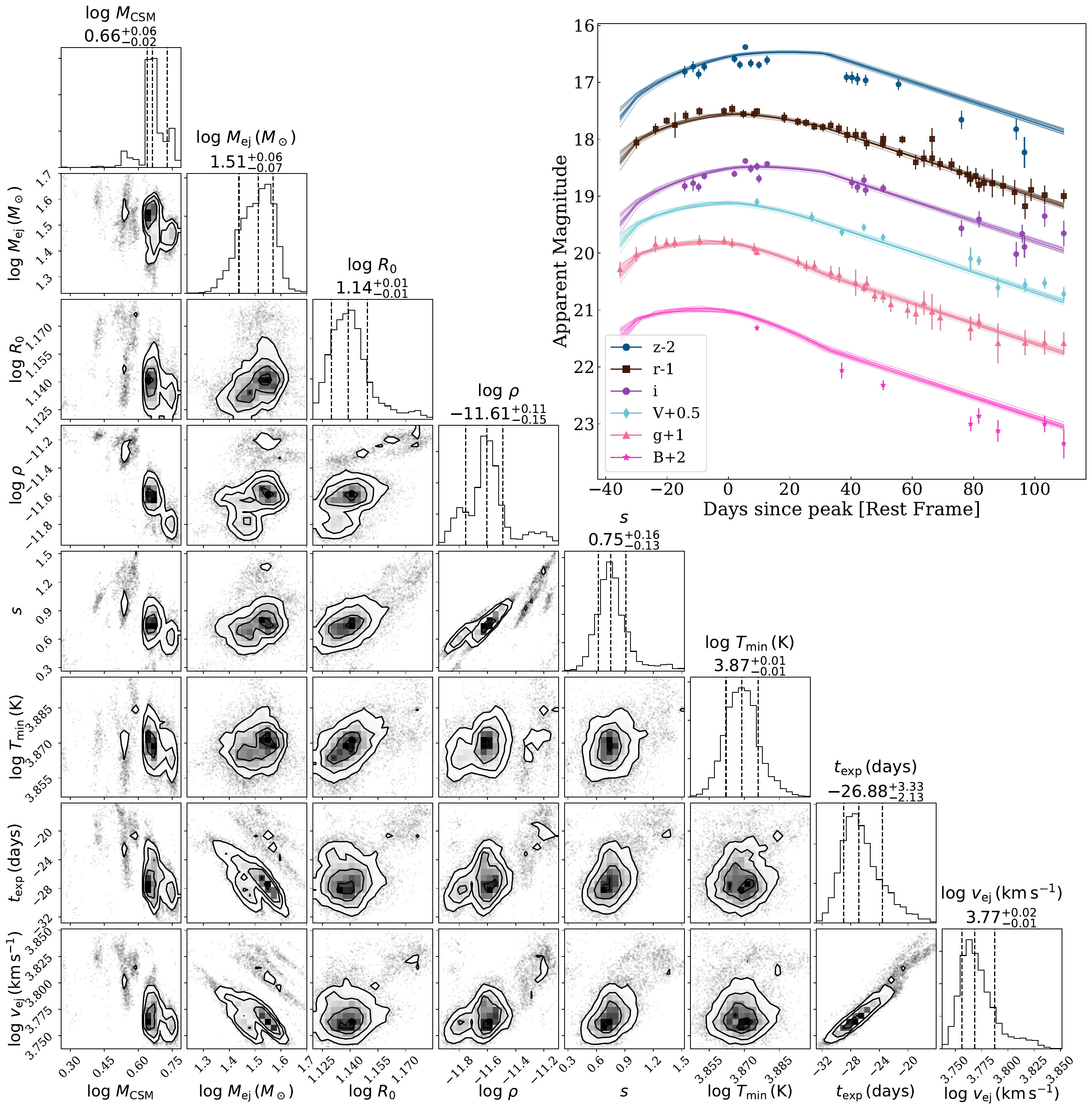}
    \caption{Corner plot of the posterior distributions and parameter correlations for the \texttt{CSM} model of SN~2019cqc. The inset shows the corresponding light-curve fit. \vspace{6mm}}
    \label{fig:csm_corner_lc}
\end{figure*}

\begin{figure*}   
    \centering
    \includegraphics[width=\textwidth]{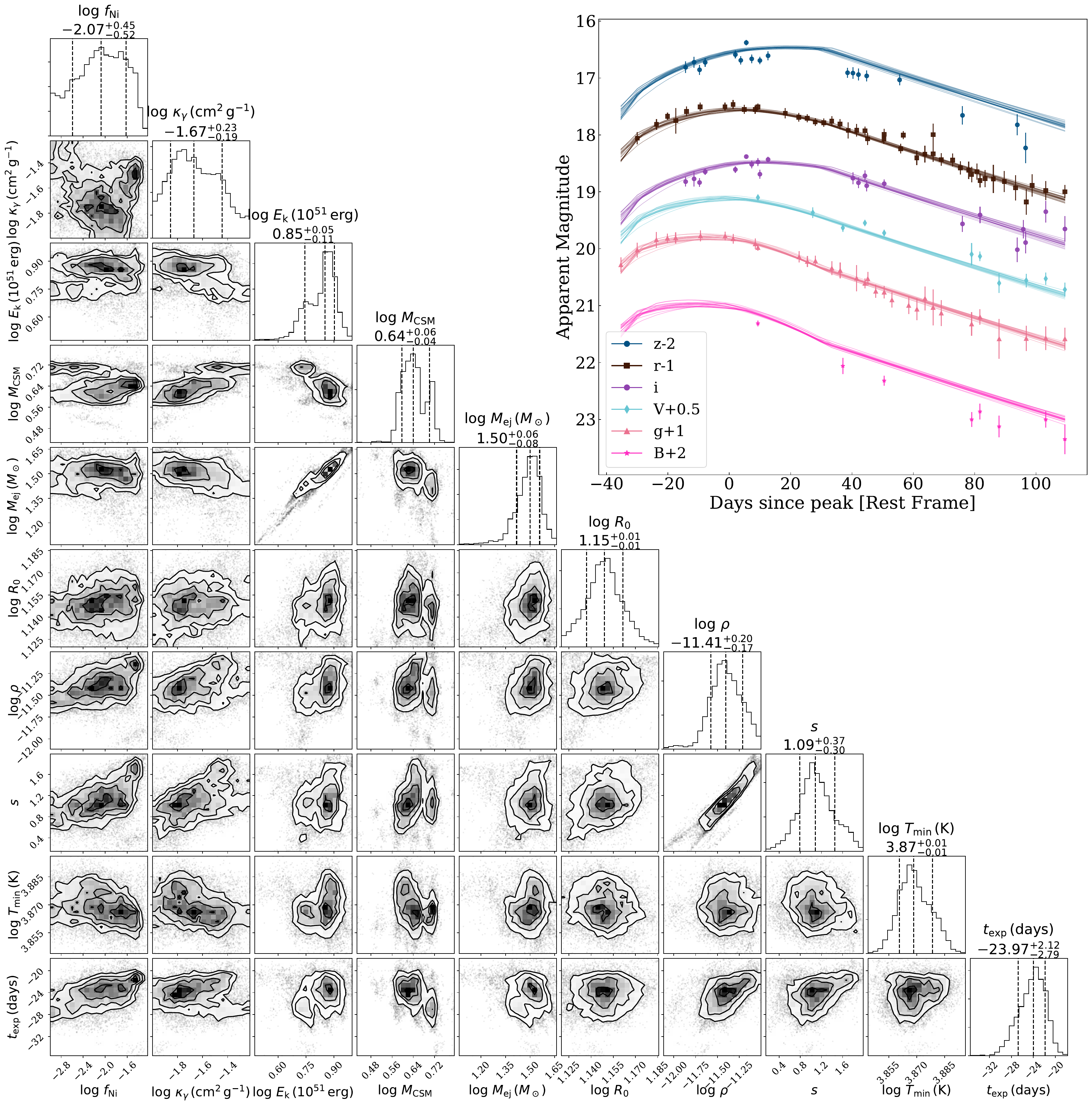}
    \caption{Corner plot of the posterior distributions and parameter correlations for the \texttt{CSMNI} model of SN~2019cqc. The inset shows the corresponding light-curve fit.\vspace{7mm}}
    \label{fig:csmni_corner_lc}   
\end{figure*}

The power-law index of the outer ejecta density profile was fixed to $n=12$, as adopted for red supergiants \citep{1999ApJ...510..379M,10.1093/mnras/stt1392}. The opacity was set to $\kappa = 0.34\ \mathrm{cm^{2}\,g^{-1}}$, consistent with fully ionised hydrogen-rich material \citep{10.1111/j.1365-2966.2011.18689.x,2012ApJ...746..121C}. For the ejecta mass ($M_{\rm ej}$), we adopted a uniform prior in the range $1$-–$100~M_{\odot}$, while the CSM mass ($M_{\rm CSM}$) was allowed to vary uniformly between $1$ and $50~M_{\odot}$. The inner CSM radius, $R_{0}$, was assigned to a uniform prior from $1$ to $150~\mathrm{AU}$. For the CSM density at this radius, $\rho_{0}$, we employed a log-uniform prior spanning 
$10^{-14}$ to $10^{-10}\ \mathrm{g\,cm^{-3}}$. We did not restrict the density profile to either a shell-like ($s=0$) or wind-like ($s=2$) configuration, but instead treated $s$ as a free parameter. For \texttt{csmni} model the mass of synthesized radioactive Ni ($M_{\rm Ni}$) is an extra parameter and a uniform prior $0.1-10$\% of $M_{\rm ej}$ has been used for the fitting. We fixed the intrinsic-scatter nuisance parameter at a minimal value of $\sigma = 0.002$. We found that this does not significantly effect the inferred physical parameters. This approach is commonly adopted to account for minimal intrinsic scatter while avoiding over-parameterization \citep[e.g.,][]{2022MNRAS.516.1193K}.
All other parameters were left free during the fitting. Parameter estimation was performed using the dynamic nested sampler \textsc{dynesty} \citep{10.1093/mnras/staa278}. The best-fitting parameters are listed in Table~\ref{tab:params}. The posterior distributions for the \texttt{csm} and \texttt{csmni} models are presented in Figures~\ref{fig:csm_corner_lc} and \ref{fig:csmni_corner_lc}, respectively, and the corresponding light-curve fits are shown in insets in each of these figures.The ejecta of the SN is large in comparison to the typical Type II and SESNe such as Ib, Ic. The inferred CSM masses from our fits are also high. These indicate an explosion associated with very massive stars ($M > 40,M_{\odot}$). Such high progenitors have also been reported for other SNe of similar Type  \citep{2014MNRAS.444.2096N,2019ARA&A..57..305G, tinyanont23}.

By analyzing the lightcurve of SN~2019cqc beyond 100 days post maximum \citet{2022MNRAS.516.1193K} found that the 
the decline rate of its lightcurve is consistent with the decay rate of radioactive $^{56}$Co. However, in the present work 
both the \texttt{csm} and \texttt{csmni} models reproduce the observed multiband lightcurves reasonably well; however, the CSM interaction parameters remain essentially unchanged when nickel is included (see Table~\ref{tab:params}). The inferred $^{56}$Ni mass is consistent with zero within $2\sigma$, implying that the contribution of radioactive heating may not be very significant, rather CSM interaction is the primary power source of this SN. 

The density profiles ($s \approx 0.7$ for the \texttt{csm} model; $s \approx 1.1$ for the \texttt{csmni} model) are inconsistent with a steady wind ($s = 2$). We compute an \emph{equivalent} mass-loss rate to contextualize the extreme CSM density. Adopting the steady-wind relation $\dot{M} = 4\pi v_w \rho_0 R_0^2$ and assuming a typical LBV wind with velocity $v_w = 100\ \mathrm{km\,s^{-1}}$, we obtain $\dot{M} \approx 0.21\,M_\odot\,\mathrm{yr^{-1}}$ and $\dot{M} \approx 0.34\,M_\odot\,\mathrm{yr^{-1}}$ for the \texttt{csm} and \texttt{csmni} models, respectively. The corresponding values of the elapsed-time after the last eruption $(t_{\rm erupt})$ before the final SN explosion, $t_{\rm \mathrm{erupt}} = R_0/v_w$, are $\sim 0.65$ \& $0.67$ yr respectively. 
It is noteworthy that the best-fit MOSFiT models also yield a parameter $t_{\mathrm{exp}}$, indicating the date of explosion with respect to the epoch of the first detection. For \texttt{csm} and \texttt{csmni} models these values are respectively $\sim -26.8$ d and $-23.9$ d. Negative signs indicate that the explosion epoch is earlier than the discovery epoch.

These inferred rates are comparable to some of the SLSN-IIn progenitors, such as SN~2015da \citep{2024MNRAS.530..405S} and ASASSN-15ua \citep{2024MNRAS.529.1205D}, where values of order $\dot{M} \sim 0.4\,M_\odot\,\mathrm{yr^{-1}}$ have been 
reported.  
We also note that some of our results differ from the findings reported by \citet{2022MNRAS.516.1193K}. Therefore, possibility of degeneracy in these models may not be ruled-out.

\begin{table*}
\centering
\caption{Best-fitting parameters from MOSFiT modelling of SN\,2019cqc using the \texttt{csm} and \texttt{csmni} models. The quoted values represent the median and the 16th/84th percentile uncertainties from the posterior distributions. The derived mass-loss rate (${\dot M}$) and the time elapsed after the last eruption $(t_{erupt})$ before the final SN explosion are also tabulated in the table.}
\label{tab:params}
\renewcommand{\arraystretch}{1.6} 
\setlength{\tabcolsep}{5pt} 

\begin{tabular}{lccccccccc}
\hline
\hline
Parameter & $M_{\rm CSM}$ & $M_{\rm ej}$ & $M_{\rm Ni}$ & $R_{0}$ & $v_{\rm ej}$ & $\rho$ & $s$ & $\dot{M}$ & $t_{\rm {\bf erupt}}$ \\
Unit & \textbf{($M_{\odot}$)} & ($M_{\odot}$) & ($M_{\odot}$) & (AU) & (km s$^{-1}$) & (g cm$^{-3}$) &  & ($M_{\odot}\,\mathrm{yr^{-1}}$) & (yr) \\
Prior & 1--50 & 1--100 & 0.1--10\% & 1--150 & 1000--40000 & $10^{-14}$--$10^{-10}$ & 0--2 & --- & --- \\
[0.8em]
\hline

\texttt{csm}   & $4.57^{+0.68}_{-0.21}$ & $32.35^{+4.80}_{-4.80}$ & --- & $13.80^{+0.32}_{-0.32}$ & $5888^{+277}_{-133}$ & $2.45^{+0.71}_{-0.71}\times 10^{-12}$ & $0.75^{+0.16}_{-0.13}$ & 0.21 & 0.65 \\
[1.2em]
\texttt{csmni} & $4.36^{+0.65}_{-0.37}$ & $31.62^{+4.68}_{-5.31}$ & $0.27^{+0.49}_{-0.19}$ & $14.12^{+0.33}_{-0.31}$ & $4750^{+770}_{-840}$ & $3.89^{+2.28}_{-1.26}\times 10^{-12}$ & $1.09^{+0.37}_{-0.30}$ & 0.34 & 0.67 \\
[0.8em]
\hline
\end{tabular}
\end{table*}

\subsection{Early evolution of H$\alpha$ Line Profile}
\label{Early Line Profile}

\begin{figure}
    \centering
    \includegraphics[width=1\linewidth]{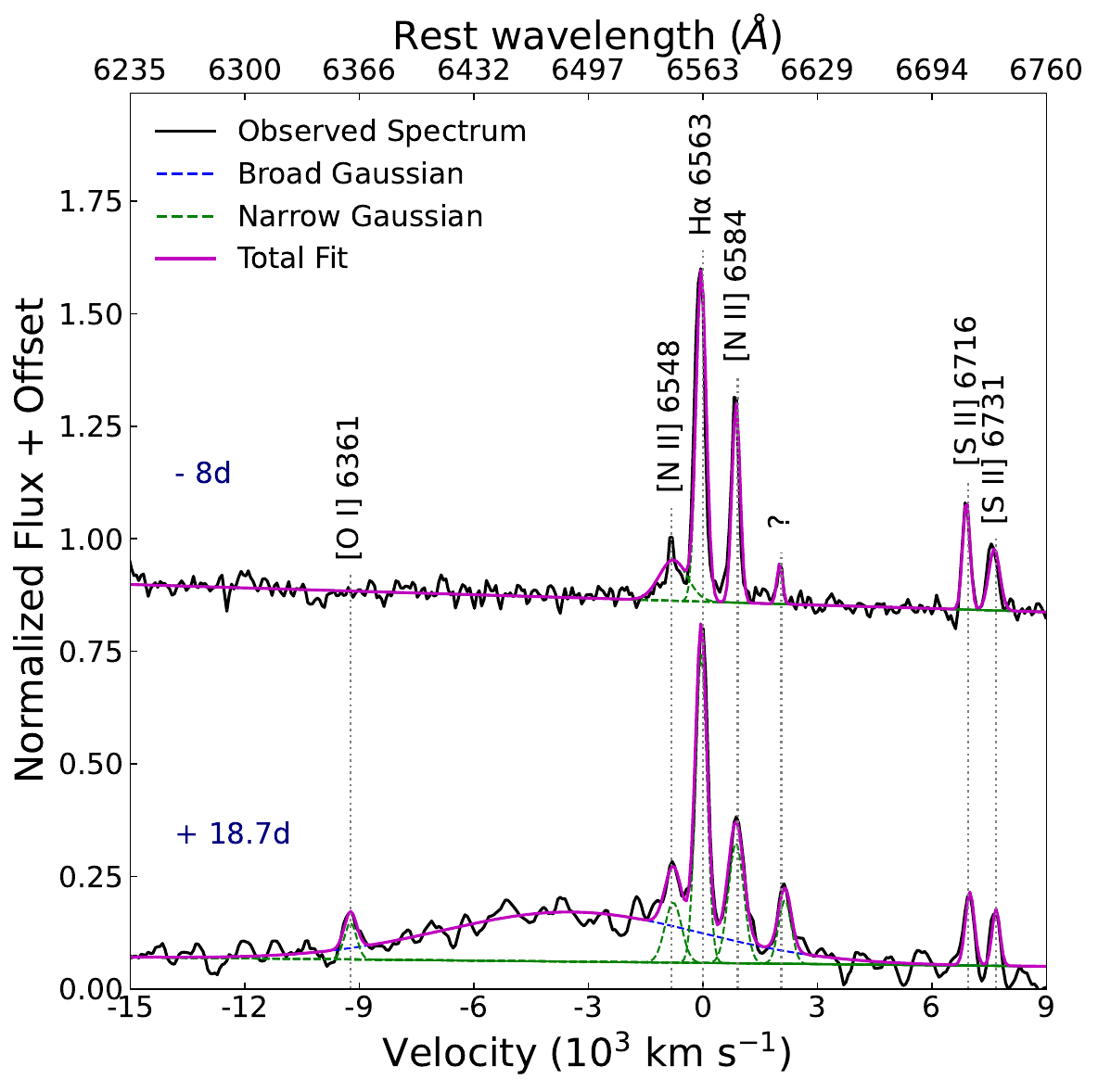}
    \caption{The figure shows the $-8$ day and $+18$ day  rest-frame spectra obtained with the P200 telescope. Prominent spectral features are identified. The key transformation is the emergence of a broad H$\alpha$ component at $+18$ days, indicative of the underlying supernova ejecta becoming visible as the optically thick CSM expands and thins. }
    \label{fig:p200}
\end{figure}

We have deblended the spectral region around H$\alpha$ using multiple Gaussian profiles with broad, intermediate, and narrow FWHM that mimic the velocity distribution in different line-forming regions along the line-of-sight (See Figure \ref{fig:p200}). The observed narrow component of the H$\alpha$ line did not show any evolution between these epochs. The FWHM remained $\sim 300 ~\mathrm{km\,s^{-1}}$. Therefore, with the existing medium/low resolution spectra, we are attributing the narrow Gaussian component in our fitting to the emission line of the host galaxy in this entire work.

The spectroscopic evolution reveals clear signatures of progressive CSM interaction. At early epochs (e.g., $-8$d), the H$\alpha$ emission profile is well described by a single, symmetric Gaussian component, only due to the emission line of the host galaxy. 
A key transformation occurs by $\sim +18$d, as the CSM becomes more transparent; a broad component emerges revealing the underlying high-velocity SN ejecta interacting with CSM.

Initially, the observed luminosity is dominated by the diffusion of shock-deposited energy through the optically thick CSM. As the interaction region expands and its optical depth decreases, emission from the shocked gas in the ejecta-CSM interaction region becomes visible, giving rise to the broad emission component. The resulting broad component shows a clearly asymmetric line profile, characterized by a prominent blue wing and an apparent deficit of flux on the red side. A more detailed discussion of the asymmetry and its temporal evolution are presented in the next section.

\begin{figure*}
    \centering
    \subfigure{%
    \includegraphics[width=0.475\textwidth]{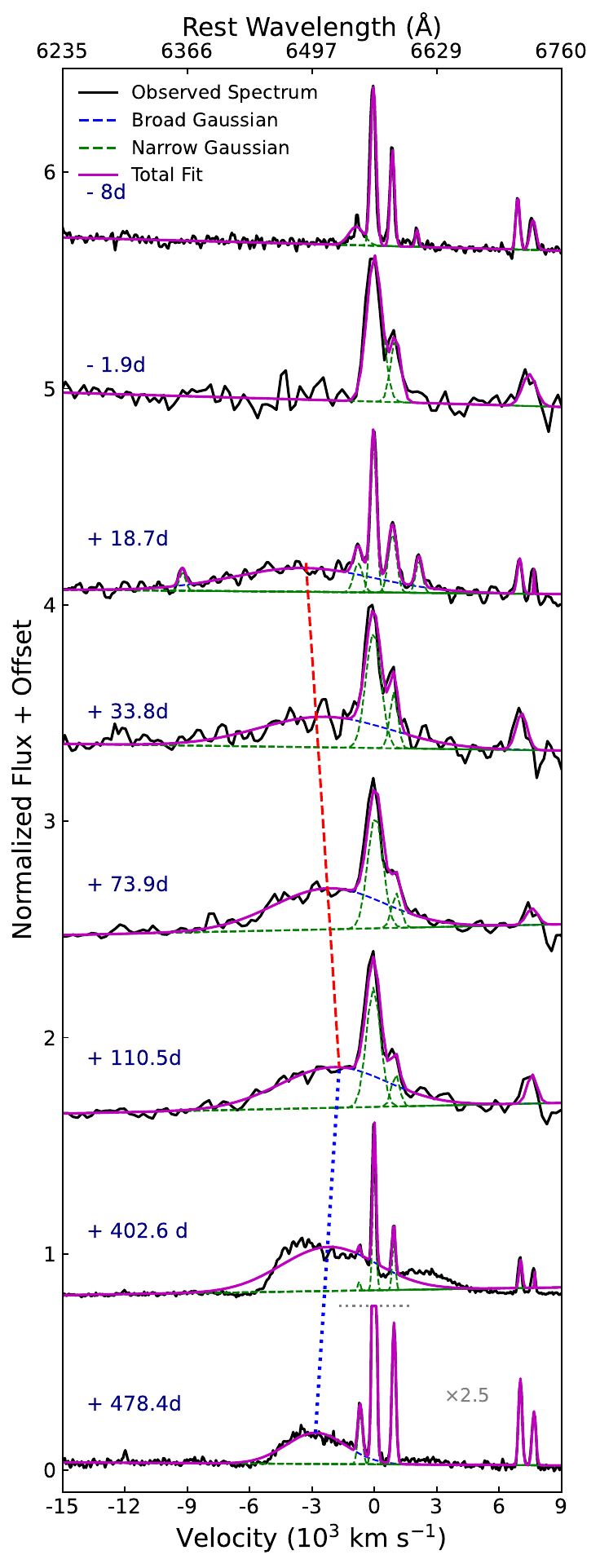}
        \label{fig:var-centre}}
    \hspace{0.004\linewidth}
    \subfigure{%
    \includegraphics[width=0.475\textwidth]{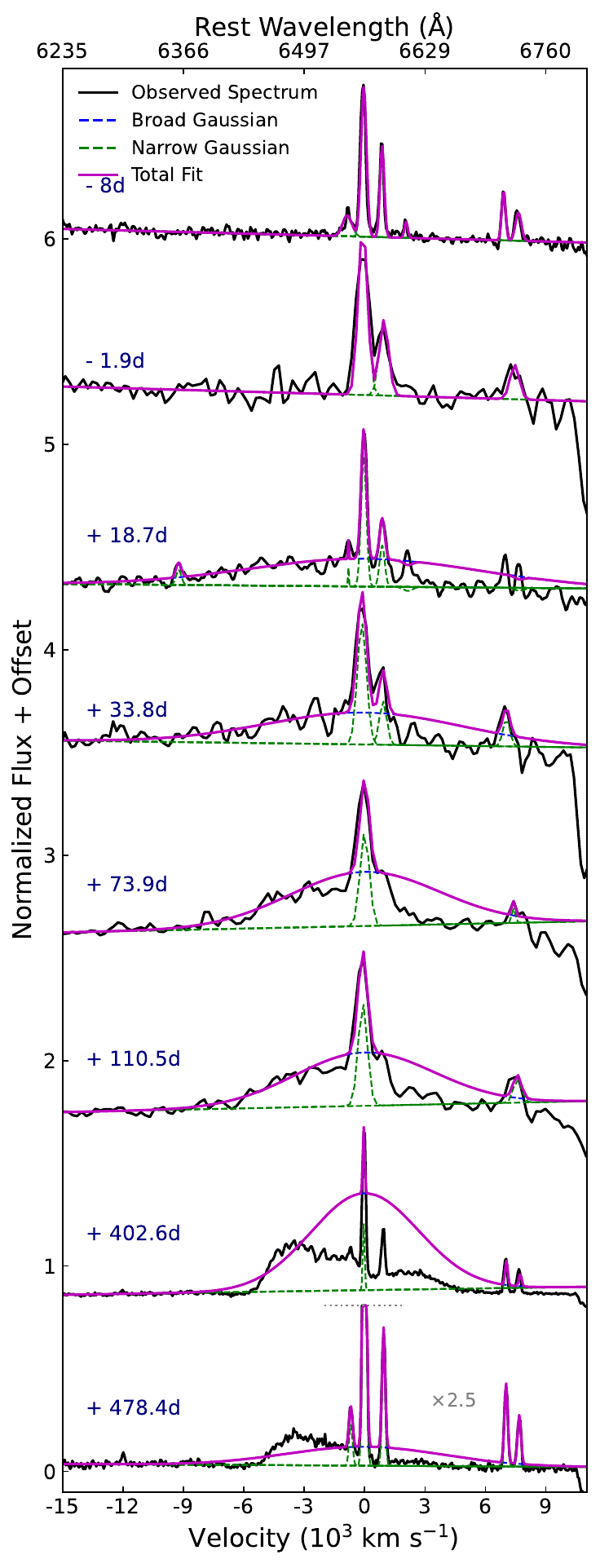}
        \label{fig:fix-centre}}
    \caption{Evolution of the H$\alpha$ profile in SN~2019cqc at selected epochs, fitted with Gaussian components. 
\textbf{Left panel:} Fits where the center and FWHM of the broad Gaussian component are allowed to vary. The peak of the broad emission progressively shifts toward the rest wavelength of H$\alpha$ until +110\,d (red dotted line), followed by a blueshift at later epochs (blue dotted line). \textbf{Right panel:} Fits where the broad Gaussian is fixed at the H$\alpha$ rest velocity. The fit uses only the blue wing; a flux deficit is consistently seen in the red wing across all epochs. The +478.4\,d spectrum is multiplied by a factor of 2.5, and its narrow Gaussians are truncated for visual clarity.}
    \label{fig:velocity-evolution}
\end{figure*}


\subsection{H$\alpha$ Line Profile Decomposition and Asymmetry}
\label{line profile decomposition}

Figure~\ref{fig:velocity-evolution} illustrates the evolution of the H$\alpha$ line profile in SN~2019cqc from early to late epochs.   
The emerging broad component is persistently blueshifted, exhibiting a prominent blue wing and relative to the rest wavelength of H$\alpha$, the overall line profile appears asymmetric. In this section, we interpret this spectral evolution within the framework of CSM interaction and evaluate the physical mechanisms that could produce the observed asymmetry.

\subsubsection{Evolution of the H$\alpha$ Profile and CSM Interaction}
\label{Evolution of the H profile}

The early-time spectra (before $+18$d) are dominated by narrow emission lines of the host galaxy, whereas 
the underlying black body continuum is produced by the optically thick ejecta.  
For our analysis, the local continuum is approximated as a straight line connecting the median flux in the intervals 6210--6300\,\AA\ (blue side) and 6640--6670\,\AA\ (red side). 
As discussed above, the narrow H$\alpha$ component of the host galaxy can be well-fitted by a symmetric Gaussian around the H$\alpha$ rest position with FWHM $\sim300~\mathrm{km\,s^{-1}}$ for the moderate resolution spectra from P200 \& MMT; whereas it is $\sim750~\mathrm{km\,s^{-1}}$ for low resolution spectra from NTT, likely indicating line-blending as well. 

The emergence of a broad H$\alpha$ component by $+18$d marks a pivotal change, revealing the underlying high-velocity SN ejecta as the optically thick CSM expands and becomes transparent. The broad emission component extends to velocities of approximately $-9000~\mathrm{km\,s^{-1}}$ on the blue side (where the negative sign shows the blueshift with respect to the rest position of H$\alpha$). The subsequent spectral evolution is defined by the growing dominance of this blueshifted broad component, while the narrow component remains unchanged. As discussed below, the velocity of the broad component decreases at later epochs,  
consistent with the expected deceleration as the forward shock sweeps up more CSM and the reverse shock propagates into the inner ejecta.

\subsubsection{Origin of the asymmetry in line profile}
The broad H$\alpha$ component in SN~2019cqc is unequivocally asymmetric, a feature observed in numerous interacting SNe. Asymmetries have been reported in various SLSNe-IIn such as SN~2017hcc \citep{2020MNRAS.499.3544S}, ASASSN-15ua \citep{2024MNRAS.529.1205D}, SN~2006tf \citep{2008ApJ...686..467S}, and SN~2015da \citep{2024MNRAS.530..405S}, as well as in the SLSN-II PS15br \citep{2018MNRAS.475.1046I}. Similar features are also seen in Type IIn SNe 2010jl \citep{2012AJ....143...17S,2014ApJ...797..118F}, KISS15s \citep{2019ApJ...872..135K}, 2007rt \citep{2009A&A...504..945T}, and 2005ip \citep{2009ApJ...695.1334S}, and even in other types, including  
interacting Type Ia events (e.g., SN~2018evt; \citealt{2023MNRAS.519.1618Y}).

\begin{figure}
    \centering
    \includegraphics[width=1\linewidth]{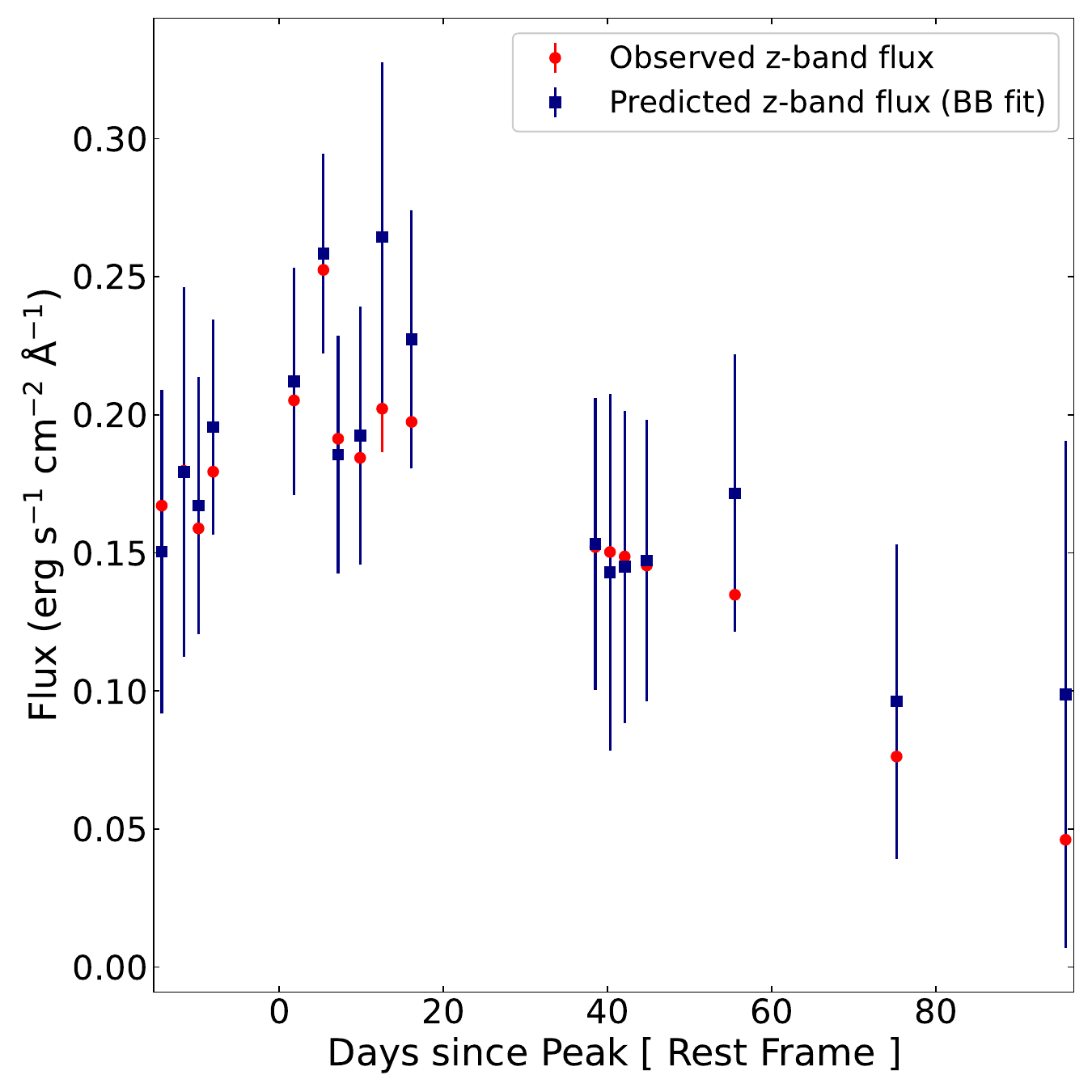}
    \caption{Assessment of infrared excess from dust. The observed \textit{z}-band flux from the LT is compared to a blackbody function extrapolated from fits to the \textit{g} and \textit{r}-band photometry. The \textit{i} band was excluded due to contamination from H$\alpha$ emission. The agreement between observed and predicted \textit{z}-band fluxes indicates no significant IR excess at these epochs, arguing against early dust formation.}
    \label{fig:z band flux}
\end{figure}

Here, we discuss different possibilities for the observed asymmetry in H$\alpha$ line profile of SN~2019cqc.\\\\ 
{\bf Dust Formation:} Newly condensed dust preferentially extinguishes light from the receding (redshifted) ejecta, leading to a blueshifted line profile, an IR excess, and a progressively more blueshifted peak over time \citep{2012AJ....143...17S}. 
Dust formation in early phases was noticed in interacting supernova SN~2006jc, which had He-rich CSM \citep{2008MNRAS.389..141M}. 
In our data, a Gaussian fit to the broad component with its center fixed at the rest wavelength of H$\alpha$ consistently overestimates the flux in the red wing (right panel of Figure~\ref{fig:velocity-evolution}) primarily mimics dust-induced suppression. However, several lines of evidence argue against dust formation as the primary cause in SN~2019cqc in these early epochs.  
In the absence of near-IR ($\textit{JHK}$) data, we compared the flux extrapolated from a blackbody fit to the \textit{g} and \textit{r} bands with the observed \textit{z}-band photometry, finding good agreement between these two (see Figure \ref{fig:z band flux} ). This suggests that either no significant dust was formed within 100 days after maximum light, or it was not significantly hot enough to produce flux excess in \textit{z}-band. Noteworthy, in SLSNe--I events, like SN~2020wnt, dust formation was observed nearly 240 days after the peak \citep{tinyanont23}. 
To investigate the possible signature of dust formation during the evolution of SN~2019cqc we searched for a mid-infrared (mid-IR) excess observed by the Near-Earth Object Wide-field Infrared Survey Explorer (NEOWISE\footnote{https://irsa.ipac.caltech.edu/data/WISE/docs/release/NEOWISE/}; \citealt{2014ApJ...792...30M}) toward the supernova. We did not notice any significant flux rise in the \textit{W1} (3.4 $\mu$m) and \textit{W2} (4.6 $\mu$m) bands of NEOWISE after the occurrence of the SN, and the measured value of flux remained constant at 15.78$\pm$0.04 mag \& 15.45$\pm$0.08 mag respectively in \textit{W1} \& \textit{W2} bands before and after the explosion. This indicates that significant dust was not formed during the evolution of SN~2019cqc, so that it can be detected with the existing sensitivity of NEOWISE.
   
Most critically, the blueshift of the broad component of SN~2019cqc does not increase with time, rather it \textit{decreases}, from $\sim -2800~\mathrm{km\,s^{-1}}$ at $+18$d to $\sim -1700~\mathrm{km\,s^{-1}}$ by $+110$d (indicated by the red line in the left panel of Figure~\ref{fig:velocity-evolution}). To get the maximum blueshift of the centroid of the broad emission component at +18d, an enormous amount of dust needs to be formed at very early epochs, which is rather unlikely. 

These facts contradict the prediction of an expanding dust shell and suggest that some other mechanisms dominate at early epochs (i.e., $\lesssim+110$d). However, in late epochs ($+402$d onward), the broad component is not symmetric and no longer exhibits a Gaussian profile. A blueshift in the peak is observed again (indicated by the blue line in the left panel of Figure~\ref{fig:velocity-evolution}), and at this epoch, the asymmetry may be caused by the formation of new dust grains in the ejecta. 
In fact, dust formation at comparable late phases was reported in Type IIn and other interaction-dominated SNe through their IR observation \citep{2004MNRAS.352..457P}.\\\\

{\bf Electron Scattering with Bulk Velocity of ejecta:}
Electron scattering in an ionized medium can also broaden and asymmetrize emission lines \citep{1948ApJ...108..116M, 2001MNRAS.326.1448C}. However, the effect is expected to diminish as the ejecta expand and the density drops, leading to increasingly symmetric profiles \citep{2012AJ....143...17S}. The persistent asymmetry in SN~2019cqc, which remains evident even at $+110$ days when the broad component originates from fast ejecta heated by the reverse shock, is inconsistent with electron scattering as the sole explanation. In objects like ASASSN-15ua and SN~2015da, the persistence of blueshifts over years also rules out a simple electron scattering origin.

\begin{figure}
    \centering
    \includegraphics[width=1\linewidth]{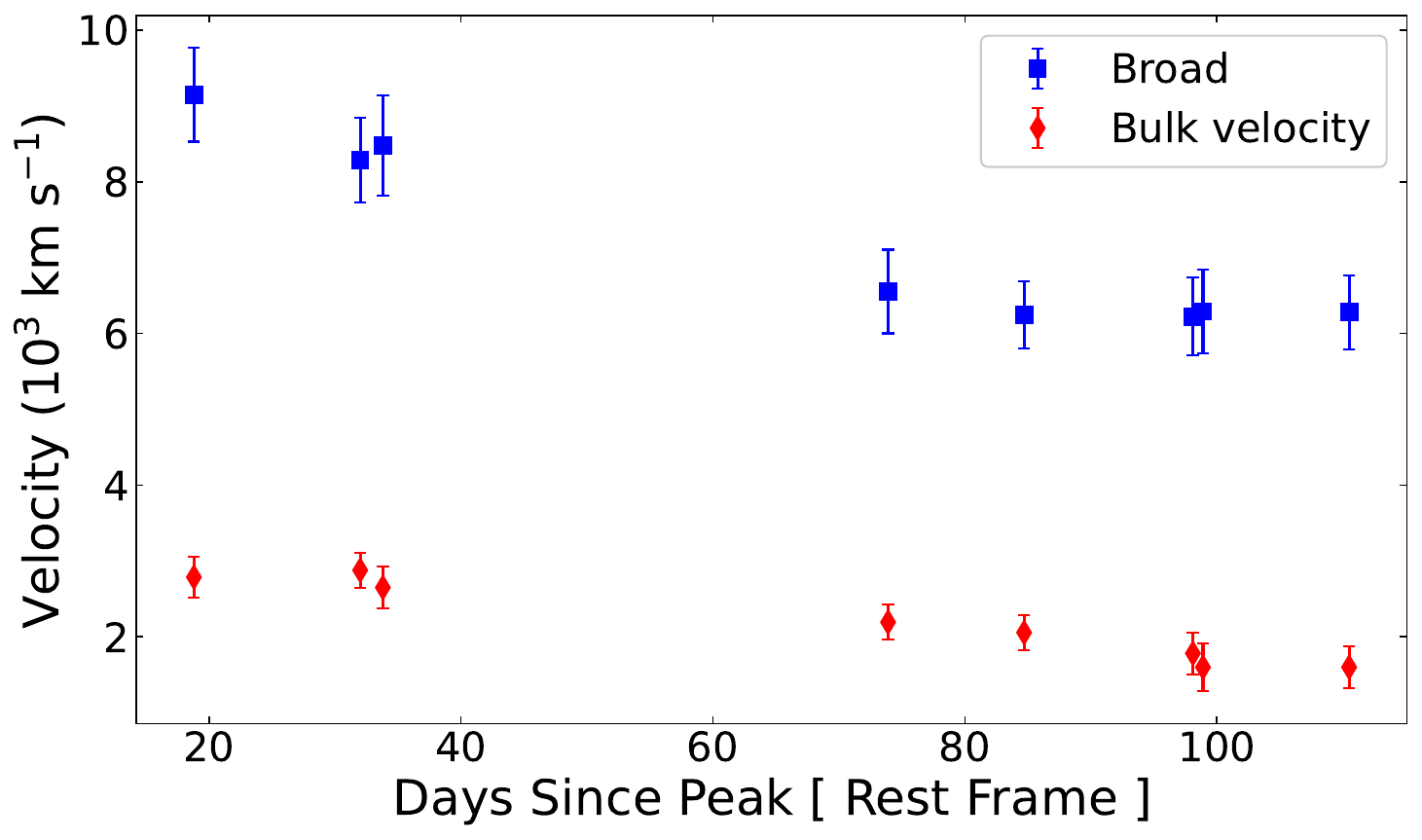}
    \caption{  Temporal evolution of the H$\alpha$ line velocity in SN~2019cqc. The plot shows the FWHM velocity of the broad Gaussian component, along with the velocity corresponding to the peak of the broad component, which appears blueshifted. The consistently blueshifted peak suggests a possible bulk motion of the emitting or scattering region, about which the broad H$\alpha$ component remains symmetric.}
    \label{fig:velocityplots}
\end{figure}

An alternative explanation is that the scattering medium (shocked-CSM) itself has a bulk velocity relative to the observer \citep{2014ApJ...797..118F}. When we fit the broad H$\alpha$ component with a Gaussian with its centroid and FWHM as free parameters, a symmetric profile provides an excellent fit to both the blue and red wings at all early epochs (Figure~\ref{fig:velocity-evolution}, left panel). 
The temporal evolution of the FWHM and the bulk velocity (which traces the scattering medium that is primarily moving coherently toward the observer, corresponds to the peak of the broad component) of the scattering medium (see Figure \ref{fig:velocityplots} ) shows that the FWHM velocity of the broad component decreases from $\sim$9000 km s$^{-1}$ at +20d to $\sim$6000 km s$^{-1}$ at +120d, whereas, the bulk velocity evolved roughly between 3000 $-$ 1500 km s$^{-1}$ within this time duration.
The systematic blueshift of this symmetric component, which decreases over time, points to a geometric or kinematic origin $-$ basically, the lateral expansion of the ejecta and interaction region over time makes the emitting region more symmetric with respect to the observer's line of sight.

In the case of SN~2019cqc, the observed decreasing blueshift could reflect the  
changing geometry of the interaction region as the shock propagates outward. This scenario provides a self-consistent explanation for the symmetric yet blueshifted broad component and its temporal evolution, without invoking dust formation at these early epochs. However, in late phases (+402d \& +478d) again a blueshift of the broad component has been noticed, which could not be modeled adopting the previous approach (as can be seen in Figure \ref{fig:velocity-evolution}). We attribute this inconsistency to the formation of new dust in the outer ejecta.\\\\

\begin{figure}
\centering

\includegraphics[width=1\linewidth]{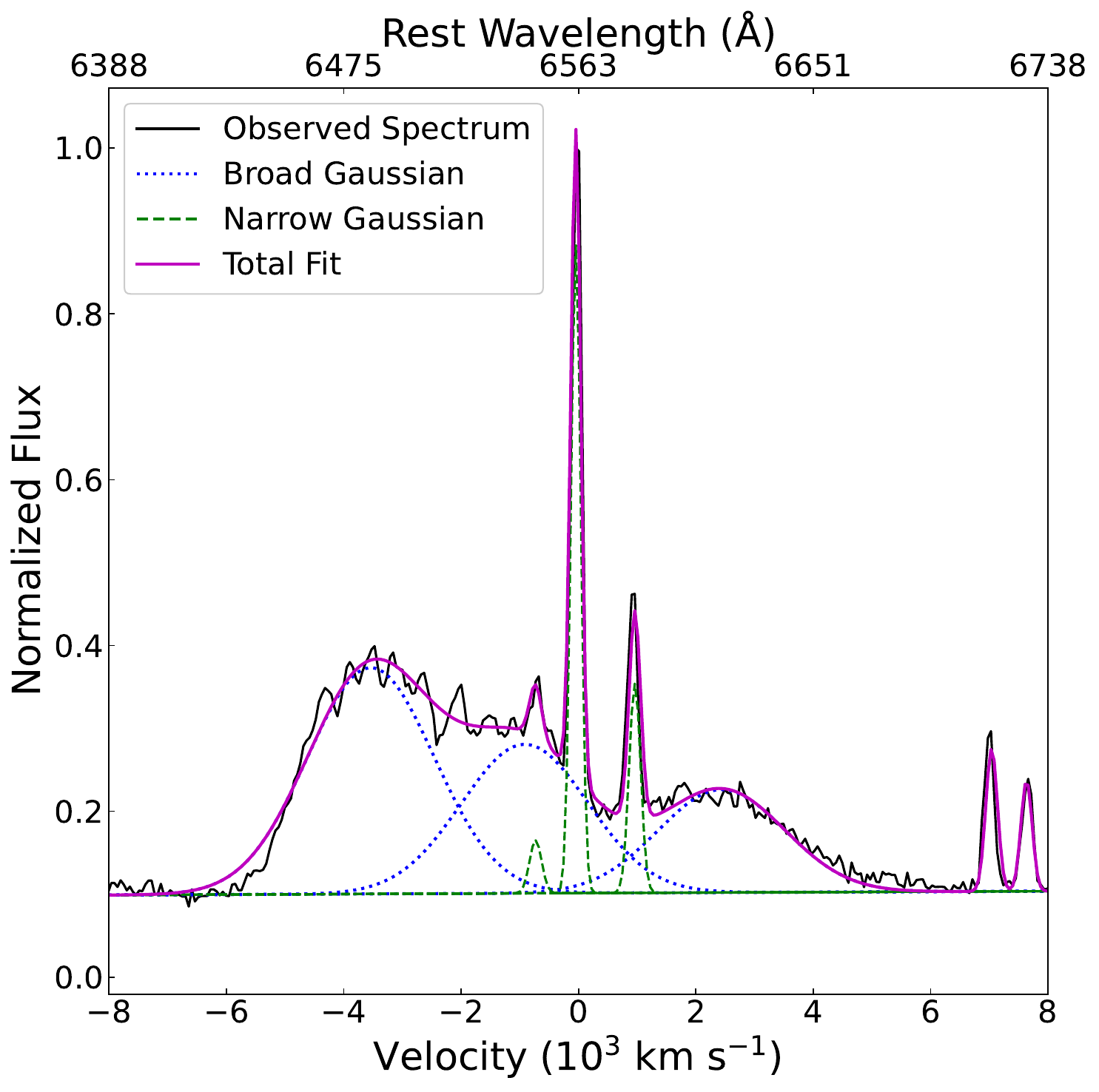}
\label{fig:mmt402}

\vspace{0.1cm}

\includegraphics[width=1\linewidth]{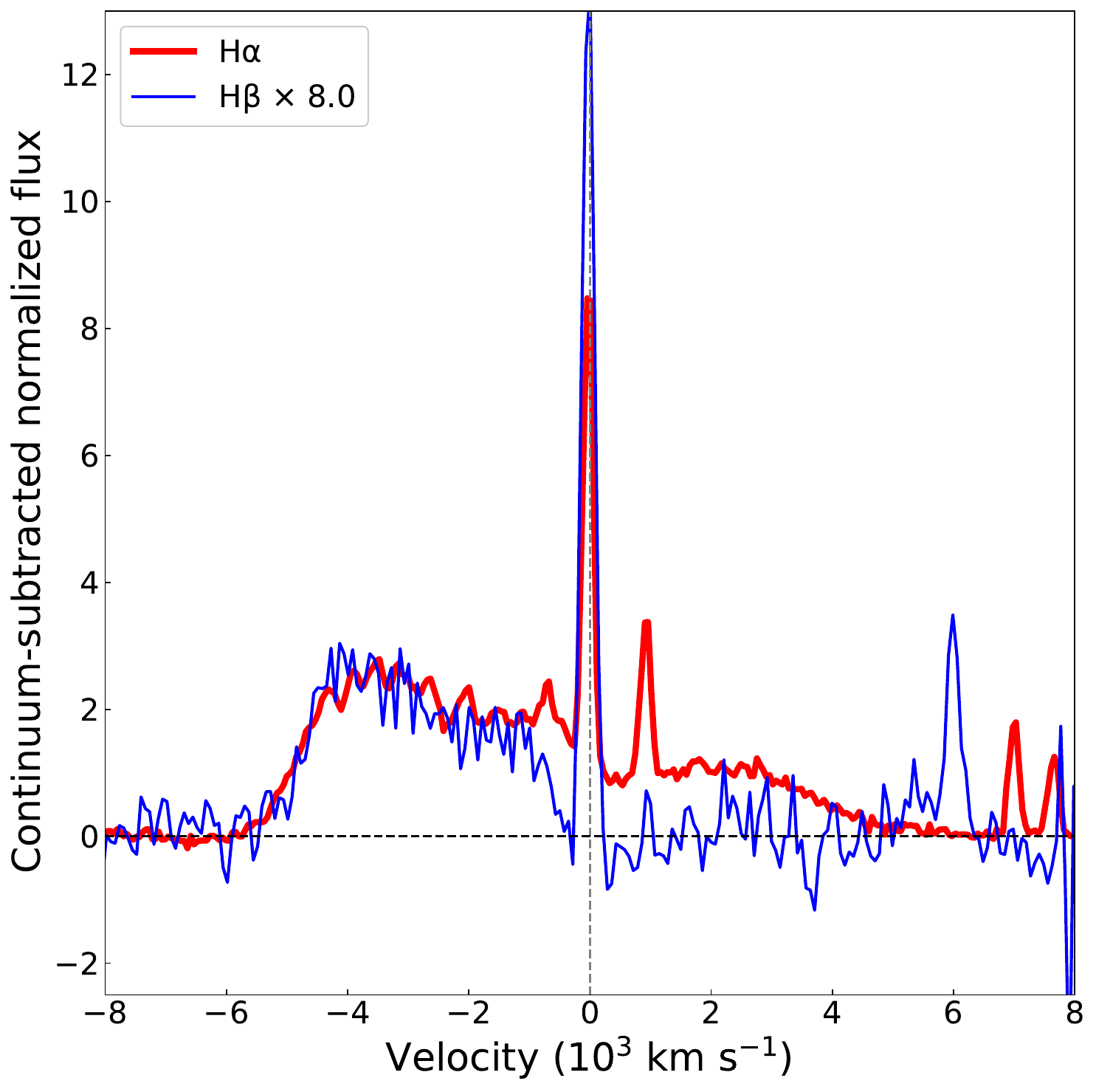}
\label{fig:halpha_hbeta}

\caption{
\textbf{Top:} Multiple-Gaussian fit to the H$\alpha$ region of the $+402$\,d MMT spectrum.
\textbf{Bottom:} Comparison between continuum-subtracted, normalized H$\alpha$ and H$\beta$ regions of the same spectrum.
}
\label{fig:halpha_analysis}
\end{figure}

{\bf Multi-component H$\alpha$ profile at late phase:} We investigated the spectrum of SN~2019cqc observed from MMT at +402 d, which has a relatively higher resolution. It shows a multi-component feature of the broad H$\alpha$ emission (see the top panel of Figure \ref{fig:halpha_analysis} ), as shown by the deblending process of this region. It indicates the presence of three probable broad Gaussians $-$ two blueshifted with velocities of centers with respect to the rest position of H$\alpha$ respectively 3540.1$\pm$37.4 km\,s$^{-1}$ \& 916.9$\pm$71.9 km\,s$^{-1}$, and one redshifted Gaussian with centroid velocity 2434.7$\pm$69.3 km\,s$^{-1}$. FWHM velocities of all three Gaussians are equal to roughly 2538 km\,s$^{-1}$. Comparison of this H$\alpha$ profile with the corresponding H$\beta$ profile (see the bottom panel of Figure \ref{fig:halpha_analysis} ) shows consistency in the positions of the blueshifted Gaussians in the velocity domain, while there is a difference in the redshifted counterpart. The average flux of H$\alpha$ profile in this region is nearly 2$\sigma$ times higher than the corresponding flux (of the continuum) in the H$\beta$ profile (where $\sigma$ is the standard deviation of the fluctuation of the continuum of the normalized H$\beta$ profile). The deficit in the flux of the red wing of the broad H$\beta$ component is mainly attributed to the line-blanketing due to Fe\,\texttt{II} multiplates present within this spectral window. The multi-component features of H$\alpha$ \& H$\beta$ at late epoch indicate the interaction of the H-rich ejecta with a ring/disc-like edge-on, and clumpy CSM $-$ advocating CSM interaction as the powering mechanism of this SN.

\vspace{10mm}

\subsection{Feature near 4600\,\AA\ Bump}
\label{feature_4600}

\begin{figure}[H]
    \centering
    \includegraphics[width=1\linewidth]{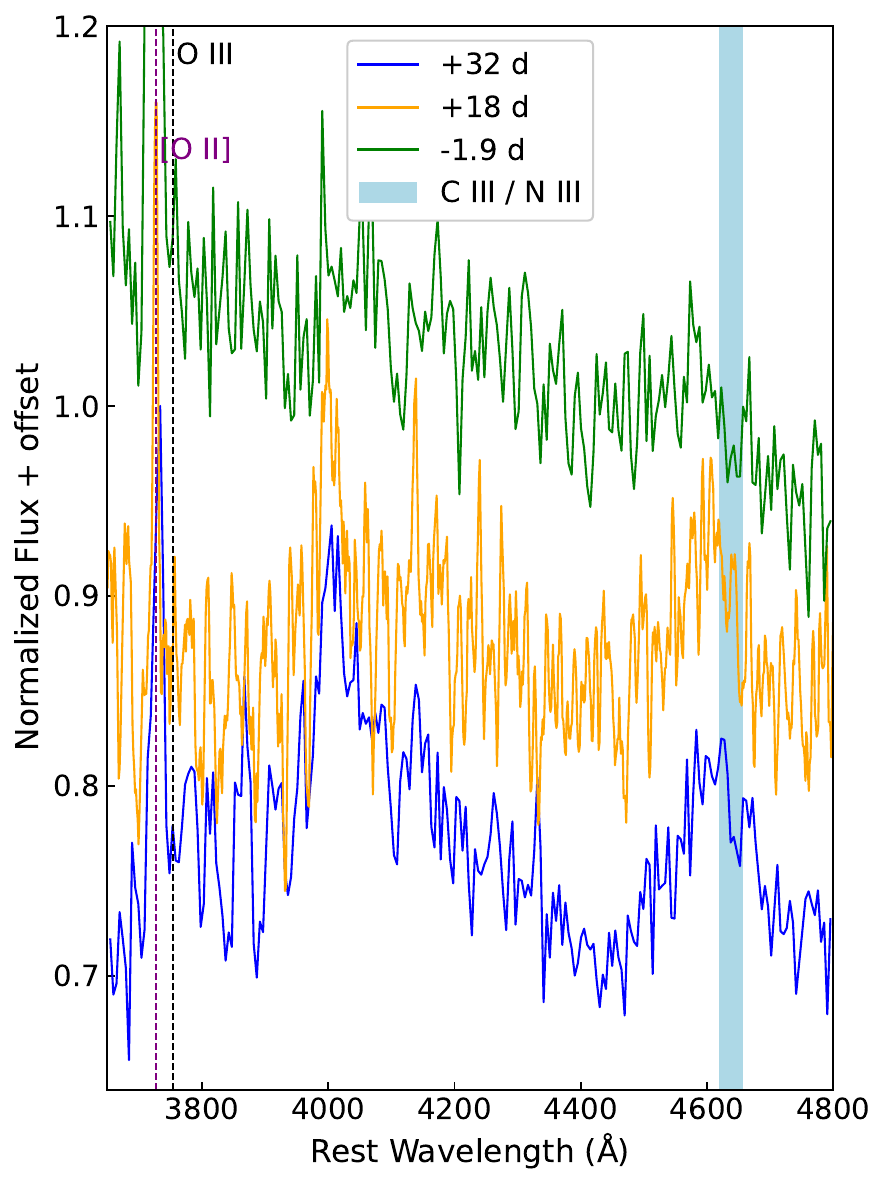}
    \caption{Normalized spectra of SN~2019cqc at three epochs relative to maximum light: $-1.9$\,d , $+18$\,d , and $+32$\,d  shown in the 3650--4800\,\AA\ region. The positions of O\,\texttt{III} $\lambda$3754 and the narrow [O\,\texttt{II}] line are marked with red dashed lines and labeled. The shaded blue region between 4640 and 4651\,\AA\ highlights the feature commonly associated with blended C\,\texttt{III} and N\,\texttt{III}. The broad excess near 4000\,\AA\ may be attributed to Fe\,\texttt{III} forest features. The spectra have been vertically offset for clarity.}
    \label{fig:bump_4600}
\end{figure}

We detect a prominent emission bump near 4600\,\AA\ in SN~2019cqc, which remains visible until +32\,d after peak (see Figure \ref{fig:bump_4600}). Similar features are frequently reported in SLSNe-II and other strongly interacting transients. In ordinary Type II supernovae, such structures are commonly attributed to flash-ionized emission \citep{2019ApJ...885...43A}; however, in the case of SN~2019cqc, any narrow, high-ionization features have not been detected, which are typically observed at very early epochs.
A comparable broad ``ledge'' near 4600\,\AA\ has been noted in normal SNe II \citep{2019ApJ...885...43A}. Several origins have been proposed. In the case of SN~2005cs, the feature was associated with high-velocity H\,$\beta$ \citep{2006MNRAS.370.1752P}, while for SN~2010id a broad, blueshifted He\,II\,$\lambda$4686 component was suggested to be the origin of this line \citep{2011ApJ...736..159G}. More recent studies, however, favor a composite of highly ionized CSM lines—-most prominently C\,III $\lambda4649$  /,  N\,III $\lambda$4634, and He\,II--rather than a single displaced transition \citep{2018MNRAS.475.1046I,2018ApJ...861...63H,2021ApJ...912...46B}

For our object, the high velocity H\,$\beta$ scenario is unlikely, since no corresponding feature is detected on the blue side of H\,$\alpha$. A pure He\,II explanation would require velocities of $\gtrsim$5000\,km\,s$^{-1}$, inconsistent with the blueshift ($\sim$2000\,km\,s$^{-1}$) measured for the broad H\,$\alpha$ emission at similar epochs. Rather, assigning the feature to blueshifted C\,III/N\,III is consistent with the velocity offset of the H\,$\alpha$ broad component, as seen in several SLSN-II events such as SN~2006tf \citep{2008ApJ...686..467S}, CSS121015 \citep{2014MNRAS.441..289B}, SN~2008es \citep{2018MNRAS.475.1046I}, PS15br, and SN~2013hx \citep{2018MNRAS.475.1046I}.

Taken together, these comparisons suggest that the 4600\,\AA\ feature in SLSNe-II is most plausibly produced by a blend of ionized C\,III/N\,III lines energized by strong interaction-powered radiation fields, rather than a single isolated transition. The persistence of this feature until around +32 d suggests the presence of very 
extended atmospheres in the progenitors of SLSNe-II events.  
Adopting the explosion epoch from the MOSFiT best-fit \texttt{csm} model 
($t_{\rm exp} = -26.9 \pm 2.7$ d) and the velocity of the broad H$\alpha$ component 
at +32 d ($v = 8200 \pm 450$ km s$^{-1}$), we infer a characteristic radius of 
$R \sim vt = (4.1 \pm 0.3)\times10^{15}$ cm for the line-forming region, indicating 
that the long-lived C\,III/N\,III signature must arise from interaction with an extended, dense circumstellar medium. The value of this radius is larger than typically observed in SNe~II, where the CSM radii are typically $10^{14}$--$10^{15}\,\mathrm{cm}$ \citep{2024ApJ...970..189J}


\section{conclusion}
\label{conclusion}
We present a detailed photometric and spectroscopic study of SN~2019cqc, a SLSN-II 
at a redshift of $z = 0.1173$. SN~2019cqc peaked at $M_g = -20.21\pm0.07$ and emitted $\sim 1.7 \times 10^{50}\,\mathrm{erg}$ of energy in the form of radiation. 
The data set includes multi-band light curves 
obtained with LT, LCO, \textit{Swift}, ATLAS, and ZTF, as well as spectroscopic 
data from the P200, ESO--NTT, and MMT. 
The photometric modelling and analysis of the spectroscopic evolution of SN~2019cqc, particularly the signature of electron scattering, and the asymmetric line profile in the late epoch indicate that the powering mechanism of this SN is CSM-interaction.
The main conclusions are summarized below.

    SN2019cqc showed a marginal bump at the late phase of its lightcurve. Multiband lightcurve modeling indicates a CSM-driven explosion. We modeled the lightcurve using MOSFiT with both the \texttt{csm} interaction model and the \texttt{csmni} model, finding that in each case CSM interaction dominates the power source.
    For the CSM model, we infer a CSM mass of $4.57^{+0.68}_{-0.21}\,M_\odot$ and an ejecta mass of $32.35^{+4.80}_{-4.80}\,M_\odot$. 
    CSM interaction is the dominant power source in the CSM+Ni model, we infer a CSM mass of $4.36^{+0.65}_{-0.37}\,M_\odot$, an ejecta mass of $31.62^{+4.68}_{-5.31}\,M_\odot$, and a $^{56}$Ni mass of $0.27^{+0.49}_{-0.19}\,M_\odot$.  
   In both models, a high mass-loss rate 
   ($\sim 0.4\,M_\odot\,\mathrm{yr}^{-1}$), together with the bumpy light curve, suggests that the progenitor experienced an LBV-like giant eruption before the SN explosion.
 
    While analyzing the spectral features, we found that at early times the H$\alpha$ profile is well described by a single, symmetric narrow Gaussian. By $\sim+18$\,d, a broad component appears with a strong blue wing, and the overall profile of the line appears asymmetric with respect to the rest wavelength. 
    Analysis of the temporal evolution of the spectra shows that the electron scattering is a dominant phenomenon in SN~2019cqc.  
    Since the broad component is symmetric about a velocity-shifted center at early times and no excess in IR flux was detected in the $z$-band, we attribute the early-time asymmetry to electron scattering, whereas the strongly non-Gaussian profile at a very late-time suggests the formation of new dust in the ejecta at late phase.

    The 4600\,\AA\ feature in SLSNe-II is likely a blend of ionized C \& N lines powered by sustained CSM interaction. Its persistence to +32\,d suggests prolonged interaction with an optically thick CSM rather than a brief flash, although a larger sample is needed to confirm this interpretation.
    
\section*{Data Availability}
 All data used in this work will be made publicly available via the WISeREP archive.
 
\begin{acknowledgments}
This study is based on observations obtained with the 200-inch Hale Telescope (P200) at Palomar Observatory.  Additional spectra were obtained at the MMT Observatory, a joint facility of the Smithsonian Institution and the University of Arizona. This work makes use of observations from the Las Cumbres Observatory network. The LCO team is supported by NSF grants AST-2308113 and AST-1911151.
 Based on observations collected at the European Organisation for Astronomical Research in the Southern Hemisphere, Chile, as part of ePESSTO+ (the advanced Public ESO Spectroscopic Survey for Transient Objects Survey). ePESSTO+ observations were obtained under ESO program IDs 1103.D-0328 and 106.216C (PI: Inserra).
We acknowledge the use of data from the Liverpool Telescope. The LT is operated on the island of La Palma by Liverpool John Moores University in the Spanish Observatorio del Roque de los Muchachos of the Instituto de Astrofisica de Canarias with financial support from the UK Science and Technology Facilities Council. We acknowledge the use of public data from the Swift data archive. This work has used the NASA Astrophysics Data System and the NASA/IPAC Extragalactic Database (NED) which is operated by Jet Propulsion Laboratory, California Institute of Technology, under contract with the National Aeronautics and Space Administration. This research has used data obtained from Pan-STARRS1 Surveys (PS1). The PS1 and the PS1 public science archive have been made possible through contributions by the Institute for Astronomy, the University of Hawaii, the Pan-STARRS Project Office, the Max-Planck Society and its participating institutes, the Max Planck Institute for Astronomy, Heidelberg and the Max Planck Institute for Extraterrestrial Physics, Garching, The Johns Hopkins University, Durham University, the University of Edinburgh, the Queen's University Belfast, the Harvard-Smithsonian Center for Astrophysics, the Las Cumbres Observatory Global Telescope Network Incorporated, the National Central University of Taiwan, the Space Telescope Science Institute, the National Aeronautics and Space Administration under Grant No. NNX08AR22G issued through the Planetary Science Division of the NASA Science Mission Directorate, the National Science Foundation Grant No. AST–1238877, the University of Maryland, Eotvos Lorand University (ELTE), the Los Alamos National Laboratory, and the Gordon and Betty Moore Foundation.This research has made use of the TNS. TNS is supported by funding from the Weizmann Institute of Science, as well as grants from the Israeli Institute for Advanced Studies and the European Union via ERC grant No. 725161. We acknowledge the use of ZTF and Lasair in this research. Lasair is supported by the UKRI Science and Technology Facilities Council and is a collaboration between the University of Edinburgh (grant ST/N002512/1) and Queen’s University Belfast (grant ST/N002520/1) within the LSST:UK Science Consortium. ZTF is supported by National Science Foundation grant AST-1440341 and a collaboration including Caltech, IPAC, the Weizmann Institute for Science, the Oskar Klein Center at Stockholm University, the University of Maryland, the University of Washington, Deutsches Elektronen-Synchrotron and Humboldt University, Los Alamos National Laboratories, the TANGO Consortium of Taiwan, the University of Wisconsin at Milwaukee, and Lawrence Berkeley National Laboratories. Operations are conducted by COO, IPAC, and UW. This research has made use of ``Aladin sky atlas’‘ developed at CDS, Strasbourg Observatory, France 2000A\&AS..143…33B and 2014ASPC..485..277B. This work has made use of data from the Asteroid Terrestrial-impact Last Alert System (ATLAS) project. The Asteroid Terrestrial-impact Last Alert System (ATLAS) project is primarily funded to search for near-Earth asteroids through NASA grants NN12AR55G, 80NSSC18K0284, and 80NSSC18K1575; by-products of the NEO search include images and catalogs from the survey area. This work was partially funded by Kepler/K2 grant J1944/80NSSC19K0112 and HST GO-15889 and STFC grants ST/T000198/1 and ST/S006109/1. The ATLAS science products have been made possible through the contributions of the University of Hawaii Institute for Astronomy, the Queen’s University Belfast, the Space Telescope Science Institute, the South African Astronomical Observatory, and the Millennium Institute of Astrophysics (MAS), Chile. TNS is supported by funding from the Weizmann Institute of Science, as well as grants from the Israeli Institute for Advanced Studies and the European Union via ERC grant No. 725161. J.R.F. is supported by the U.S. National Science Foundation (NSF) Graduate Research Fellowship Program under grant 2139319. MN is supported by the European Research Council (ERC) under the European Union’s Horizon 2020 research and innovation programme (grant agreement No.~948381). 
DH acknowledges Manipal Centre for Natural Sciences (MCNS), Manipal Academy of Higher Education (MAHE) for facilities and support. RR acknowledges the Institute of Astronomy Space and Earth Science (IASES) for facilities and support. RR is the visiting Associate of the Inter-University Centre for Astronomy and Astrophysics (IUCAA), Pune, India. 
TK acknowledges support from the Research Council of Finland project 360274. T.-W.C. acknowledges financial support from the Yushan Fellow Program of the Ministry of Education, Taiwan (MOE-111-YSFMS-0008-001-P1), and from the National Science and Technology Council, Taiwan (NSTC 114-2112-M-008-021-MY3). SM acknowledges financial support from the Research Council of Finland project 350458.

\end{acknowledgments}

\bibliography{sample701}{}
\bibliographystyle{aasjournalv7}

\appendix
\section{}

The standard stars used for calibration are listed in Table~\ref{tab:refstar}. The calibrated magnitudes of SN2019cqc obtained from LCO, LT, and Swift observations are listed in Tables~\ref{tab:lco_photometry}, \ref{tab:lt_photometry}, and \ref{tab:swift_19cqc}, respectively. The journal of spectroscopic observations is provided in Table~\ref{tab:obslog}.

\FloatBarrier
\restartappendixnumbering


\begin{table*}
\centering
\caption{Identification number (ID), coordinates and calibrated magnitudes of stable secondary standard stars in the field of SN2019cqc. SDSS magnitudes were derived by converting Pan-STARRS1 magnitudes following \citet{2012ApJ...750...99T}}
\label{tab:refstar}
\begin{tabular}{ccccccc}
\hline
ID & RA (J2000) & Dec (J2000) & $g_{\rm SDSS}$ & $r_{\rm SDSS}$ & $i_{\rm SDSS}$ & $z_{\rm SDSS}$ \\
 & (deg) & (deg) & (AB mag) & (AB mag) & (AB mag) & (AB mag) \\
\hline
1 & 275.4240 & 30.9872 & 19.864 $\pm$ 0.013 & 19.184 $\pm$ 0.010 & 18.925 $\pm$ 0.007 & 18.826 $\pm$ 0.014 \\
2 & 275.4269 & 30.9879 & 17.160 $\pm$ 0.010 & 16.573 $\pm$ 0.006 & 16.393 $\pm$ 0.005 & 16.311 $\pm$ 0.012 \\
3 & 275.4474 & 30.9957 & 17.614 $\pm$ 0.010 & 17.072 $\pm$ 0.008 & 16.842 $\pm$ 0.005 & 16.783 $\pm$ 0.012 \\
4 & 275.4436 & 31.0052 & 17.637 $\pm$ 0.011 & 16.865 $\pm$ 0.007 & 16.597 $\pm$ 0.006 & 16.452 $\pm$ 0.013 \\
5 & 275.4373 & 31.0045 & 20.732 $\pm$ 0.024 & 19.728 $\pm$ 0.014 & 19.392 $\pm$ 0.011 & 19.161 $\pm$ 0.024 \\
6 & 275.4226 & 30.9971 & 20.767 $\pm$ 0.028 & 19.728 $\pm$ 0.014 & 19.313 $\pm$ 0.014 & 19.079 $\pm$ 0.025 \\
7 & 275.4087 & 30.9878 & 17.245 $\pm$ 0.009 & 16.692 $\pm$ 0.005 & 16.492 $\pm$ 0.005 & 16.412 $\pm$ 0.019 \\
8 & 275.4445 & 30.9741 & 17.381 $\pm$ 0.009 & 16.862 $\pm$ 0.008 & 16.702 $\pm$ 0.005 & 16.644 $\pm$ 0.012 \\
9 & 275.4248 & 30.0049 & 22.393 $\pm$ 0.095 & 21.191 $\pm$ 0.044 & 20.667 $\pm$ 0.020 & 20.350 $\pm$ 0.034 \\
10 & 275.4120 & 30.9747 & 18.518 $\pm$ 0.015 & 17.281 $\pm$ 0.008 & 16.848 $\pm$ 0.008 & 16.608 $\pm$ 0.018 \\
\hline
\end{tabular}
\end{table*}

\begin{table*}
\centering
\caption{LCO  Photometry}
\label{tab:lco_photometry}
\begin{tabular}{cccccc}
\hline
\input{LCO_BVgri.txt}
\end{tabular}
\end{table*}

\begin{table*}
\centering
\label{tab:lt_photometry}
\caption{LT Photometry}
\label{tab:lt_photometry}
\begin{tabular}{ccccc}
\hline
\input{LT_griz.txt}
\end{tabular}
\end{table*}


\begin{table*}
\caption{Swift photometry of SN2019cqc$^a$}
\label{tab:swift_19cqc}
\begin{tabular}{c c c c c c c}
\hline
MJD & $\mathit{ uvw2}$ & $\mathit{ uvm2}$ & $\mathit{ uvw1}$ & $\mathit{ u}$ & $\mathit{b}$ & $\mathit{v}$ \\
 & (AB mag) & (AB mag) & (AB mag) & (AB mag) & (AB mag) & (AB mag) \\
\hline
58625.265 & $21.861\pm0.226$ & $21.425\pm0.169$ & $20.835\pm0.181$ & $19.715\pm0.147$ & $18.747\pm0.119$ & $18.407\pm0.172$ \\
58640.443 & $21.460\pm0.144$ & $21.770\pm0.211$ & $20.907\pm0.144$ & $19.897\pm0.112$ & \dots & \dots \\
58647.721 & $21.476\pm0.146$ & $21.724\pm0.200$ & $21.033\pm0.169$ & $20.143\pm0.134$ & \dots & \dots \\
58654.420 & $21.581\pm0.132$ & $21.647\pm0.178$ & $21.185\pm0.154$ & $20.203\pm0.116$ & \dots & \dots \\

\hline
\end{tabular}
$^a$ In this work, we have used only those {\it Swift} observations which were performed within +200 days post maximum (between MJD 58625.221 \& 58654.42). Beyond that, the object was not detected from {\it Swift}, and is not reported in this table.
\end{table*}


\begin{table*}
\caption{Journal of spectroscopic observations of SN2019cqc.}
\label{tab:obslog}
\begin{center}
\setlength{\tabcolsep}{1.4pt}
\begin{tabular}{ccccccccc}
\hline
\hline
UT Date & MJD & Phase$^{a}$ & Telescope / Instrument & Exposure & Grating & Dispersion$^{f}$ & Airmass \\
(YYYY/MM/DD/ HH:MM) & (days) &   & & (s) & ($\text{gr mm}^{-1}$) &  & \\
\hline

2019/04/24/ 11:29 & 58597.979 & -9.3 & P200  / DBSP $^{b}$ & 500 & 600/4000+316/7500$^{e}$
 & 02/72 & 2.021 \\
2019/05/01/ 08:19 & 58604.847 & -2.5 & NTT  / EFOSC2 $^{c}$ & 900 & 236 & 2.77 & 2.012 \\
2019/05/02/ 08:12 & 58605.842 & -1.5 & NTT / EFOSC2 & 2700 & 300 & 2.04 & 2.014 \\
2019/05/11/ 07:08 & 58614.798 & 7.5 & NTT / EFOSC2 & 2700 & 300 & 2.12 & 2.060 \\
2019/05/11/ 08:08 & 58614.839 & 7.5 & NTT / EFOSC2 & 2700 & 300 & 2.04 & 2.018 \\
2019/05/14/ 06:53 & 58617.787 & 10.5 & NTT / EFOSC2 & 1800 & 236 & 2.77 & 2.067 \\
2019/05/14/ 07:29 & 58617.812 & 10.5 & NTT / EFOSC2 & 1800 & 236 & 2.77 & 2.011 \\
2019/05/24/ 11:15 & 58627.969 & 20.7 & P200 / DBSP & 500 &  600/4000+316/7500 & 02/72 & 1.095 \\
2019/05/24/ 11:25 & 58627.969 & 20.7 & P200 / DBSP & 500 & 600/4000+316/7500 & 02/72 & 1.090 \\
2019/06/08/ 05:18 & 58642.721 & 35.4 & NTT / EFOSC2 & 2699 & 236 & 2.77 & 2.060 \\
2019/06/08/ 06:09 & 58642.756 & 35.5 & NTT / EFOSC2 & 2700 & 236 & 2.77 & 2.011 \\
2019/06/10/ 04:23 & 58644.683 & 37.4 & NTT / EFOSC2 & 2700 & 300 & 2.12 & 2.263 \\
2019/06/10/ 05:23 & 58644.724 & 37.4 & NTT / EFOSC2 & 2700 & 300 & 2.12 & 2.033 \\
2019/06/10/ 06:23 & 58644.766 & 37.5 & NTT / EFOSC2 & 2700 & 300 & 2.12 & 2.037 \\
2019/06/29/ 07:53  & 58663.328 & 56.0 & MMT / Binospec  $^{d}$&   600    & 270   & 1.29  & 1.009    \\
2019/07/25/ 02:00 & 58689.584 & 82.3 & NTT / EFOSC2 & 2700 & 236 & 2.77 & 2.098 \\
2019/07/25/ 02:52 & 58689.620 & 82.3 & NTT / EFOSC2 & 2700 & 236 & 2.77 & 2.009 \\
2019/08/06/ 01:27 & 58701.561 & 94.3 & NTT / EFOSC2 & 2700 & 236 & 2.77 & 2.057 \\
2019/08/06/ 02:21 & 58701.599 & 94.3 & NTT / EFOSC2 & 2700 & 236 & 2.77 & 2.014 \\
2019/08/06/ 03:18 & 58701.638 & 94.3 & NTT / EFOSC2 & 2700 & 236 & 2.77 & 2.159 \\
2019/08/21/ 02:29 & 58716.604 & 109.3 & NTT / EFOSC2 & 2700 & 236 & 2.77 & 2.209 \\
2019/08/22/ 01:24 & 58717.559 & 110.3 & NTT / EFOSC2 & 1522 & 236 & 2.77 & 2.020 \\
2019/09/04/ 00:09 & 58730.506 & 123.2 & NTT / EFOSC2 & 2700 & 236 & 2.77 & 2.010 \\
2020/07/26/ 06:53 &59056.287 & 448.9 &  MMT / Binospec   & 2790 & 270 &1.29 & 1.053 \\
2020/10/19/ 09:30 & 59141.000 & 584   & MMT / Binospec   & 600  & 270 & 1.29 &  1.053   \\
\hline
\end{tabular}
\end{center}
$^{a}$ Days relative to the peak brightness in the $r$ band (MJD 58607.30) in the observer frame.\\
$^{b}$ 200-inch Hale Telescope / Double Spectrograph. \\
$^{c}$  European Southern Observatory New Technology Telescope/ ESO Faint Object Spectrograph and Camera. \\
$^{d}$ Multiple Mirror Telescope / Binospec.\\
$^{e}$ Gratings are given as lines mm$^{-1}$/ blaze wavelength ( \AA\ ) for the blue and red arms.\\
$^{f}$ For P200, dispersion is given in \AA\ mm$^{-1}$. For EFOSC2 and Binospec, dispersion is given in \AA\ pixel$^{-1}$.
\end{table*}



\end{document}

%% file: LCO_BVgri.txt
MJD & \textit{ B} & \textit{V} & \textit{gp} & \textit{rp} & \textit{ip}\\
 & (AB mag) & (AB mag) & (AB mag) & (AB mag) & (AB mag) \\
\hline
58617.441 & 19.313 $\pm$ 0.051 & 18.6 $\pm$ 0.031 & 18.981 $\pm$ 0.057 & 18.508 $\pm$ 0.057 & 18.476 $\pm$ 0.071 \\
58625.438 & 19.534 $\pm$ 0.067 & 18.697 $\pm$ 0.038 & 19.07 $\pm$ 0.065 & 18.536 $\pm$ 0.06 & 18.453 $\pm$ 0.069 \\
58637.426 & -- & 18.867 $\pm$ 0.093 & -- & -- & -- \\
58641.345 & 19.832 $\pm$ 0.063 & 18.831 $\pm$ 0.035 & 19.322 $\pm$ 0.06 & 18.625 $\pm$ 0.057 & 18.487 $\pm$ 0.045 \\
58648.362 & 20.061 $\pm$ 0.14 & 19.133 $\pm$ 0.072 & -- & -- & -- \\
58649.429 & 20.012 $\pm$ 0.111 & 18.956 $\pm$ 0.049 & 19.423 $\pm$ 0.098 & 18.6 $\pm$ 0.041 & 18.521 $\pm$ 0.049 \\
58656.398 & 19.793 $\pm$ 0.111 & 19.048 $\pm$ 0.045 & 19.609 $\pm$ 0.087 & 18.928 $\pm$ 0.077 & 18.718 $\pm$ 0.096 \\
58663.42 & 20.319 $\pm$ 0.09 & 19.224 $\pm$ 0.056 & 19.764 $\pm$ 0.109 & 18.99 $\pm$ 0.097 & 18.858 $\pm$ 0.077 \\
58670.404 & 20.303 $\pm$ 0.084 & 19.283 $\pm$ 0.042 & 19.746 $\pm$ 0.073 & 19.003 $\pm$ 0.071 & 18.695 $\pm$ 0.069 \\
58681.273 & -- & -- & 19.92 $\pm$ 0.15 & 18.995 $\pm$ 0.192 & 18.743 $\pm$ 0.225 \\
58688.331 & 20.457 $\pm$ 0.095 & 19.372 $\pm$ 0.066 & 20.066 $\pm$ 0.106 & 19.277 $\pm$ 0.085 & 18.816 $\pm$ 0.16 \\
58695.269 & 21.001 $\pm$ 0.132 & 19.599 $\pm$ 0.197 & 20.323 $\pm$ 0.21 & 19.698 $\pm$ 0.136 & -- \\
58698.324 & 20.862 $\pm$ 0.139 & 19.634 $\pm$ 0.084 & 20.2 $\pm$ 0.135 & 19.803 $\pm$ 0.125 & 19.407 $\pm$ 0.143 \\
58705.292 & 21.124 $\pm$ 0.191 & 20.106 $\pm$ 0.174 & 20.583 $\pm$ 0.351 & -- & -- \\
58715.157 & -- & 20.059 $\pm$ 0.119 & 20.578 $\pm$ 0.221 & 20.177 $\pm$ 0.225 & -- \\
58722.272 & 21.002 $\pm$ 0.144 & 20.026 $\pm$ 0.107 & 20.572 $\pm$ 0.21 & 19.98 $\pm$ 0.169 & 19.35 $\pm$ 0.189 \\
58729.24 & 21.349 $\pm$ 0.262 & 20.219 $\pm$ 0.122 & 20.583 $\pm$ 0.195 & 19.998 $\pm$ 0.117 & 19.654 $\pm$ 0.226 \\
58753.184 & 21.342 $\pm$ 0.197 & 20.292 $\pm$ 0.137 & 20.71 $\pm$ 0.226 & 20.679 $\pm$ 0.358 & 19.874 $\pm$ 0.191 \\
58762.154 & 21.948 $\pm$ 0.393 & 20.724 $\pm$ 0.217 & 21.072 $\pm$ 0.335 & -- & 19.842 $\pm$ 0.273 \\
58769.134 & -- & 20.169 $\pm$ 0.172 & 20.614 $\pm$ 0.265 & 19.799 $\pm$ 0.183 & 19.375 $\pm$ 0.223 \\
58779.099 & 21.342 $\pm$ 0.175 & 20.175 $\pm$ 0.104 & 20.575 $\pm$ 0.168 & 19.925 $\pm$ 0.141 & 19.547 $\pm$ 0.215 \\
58789.103 & 21.019 $\pm$ 0.145 & 19.969 $\pm$ 0.116 & 20.666 $\pm$ 0.161 & 19.663 $\pm$ 0.107 & 19.444 $\pm$ 0.13 \\
58802.059 & 21.258 $\pm$ 0.217 & -- & -- & -- & -- \\
58807.049 & 21.281 $\pm$ 0.237 & 20.191 $\pm$ 0.125 & 20.722 $\pm$ 0.187 & -- & 19.839 $\pm$ 0.184 \\
\hline

%% file: LT_griz.txt
MJD & \textit{g} & \textit{r} & \textit{i} & \textit{z}\\
& (AB mag) & (AB mag) & (AB mag) & (AB mag) \\
\hline
58591.101 & 18.752 $\pm$ 0.073 & 18.764 $\pm$ 0.061 & 18.822 $\pm$ 0.086 & 18.811 $\pm$ 0.102 \\
58594.181 & 18.819 $\pm$ 0.07 & 18.655 $\pm$ 0.053 & 18.772 $\pm$ 0.134 & 18.724 $\pm$ 0.096 \\
58596.167 & 18.836 $\pm$ 0.053 & 18.649 $\pm$ 0.052 & 18.837 $\pm$ 0.077 & 18.856 $\pm$ 0.089 \\
58598.223 & 18.85 $\pm$ 0.04 & 18.608 $\pm$ 0.042 & 18.648 $\pm$ 0.056 & 18.725 $\pm$ 0.077 \\
58609.214 & 18.879 $\pm$ 0.038 & 18.483 $\pm$ 0.044 & 18.609 $\pm$ 0.057 & 18.589 $\pm$ 0.068 \\
58611.185 & 18.915 $\pm$ 0.041 & 18.515 $\pm$ 0.056 & -- & 18.69 $\pm$ 0.074 \\
58613.189 & 18.859 $\pm$ 0.035 & 18.46 $\pm$ 0.033 & 18.382 $\pm$ 0.034 & 18.379 $\pm$ 0.05 \\
58615.166 & 18.973 $\pm$ 0.042 & 18.721 $\pm$ 0.057 & 18.518 $\pm$ 0.069 & 18.664 $\pm$ 0.075 \\
58618.145 & 19.045 $\pm$ 0.051 & 18.704 $\pm$ 0.047 & 18.691 $\pm$ 0.078 & 18.693 $\pm$ 0.068 \\
58621.127 & 19.173 $\pm$ 0.083 & 18.569 $\pm$ 0.054 & 18.433 $\pm$ 0.049 & 18.609 $\pm$ 0.081 \\
58625.111 & 19.113 $\pm$ 0.056 & 18.61 $\pm$ 0.044 & 18.583 $\pm$ 0.053 & 18.628 $\pm$ 0.077 \\
58650.085 & 19.581 $\pm$ 0.112 & 18.947 $\pm$ 0.077 & 18.649 $\pm$ 0.078 & 18.908 $\pm$ 0.089 \\
58652.032 & 19.524 $\pm$ 0.157 & 18.97 $\pm$ 0.095 & 18.761 $\pm$ 0.104 & 18.914 $\pm$ 0.103 \\
58654.052 & 19.961 $\pm$ 0.126 & 19.109 $\pm$ 0.093 & 18.841 $\pm$ 0.095 & 18.94 $\pm$ 0.105 \\
58657.046 & 19.813 $\pm$ 0.088 & 19.119 $\pm$ 0.081 & 18.893 $\pm$ 0.101 & 18.963 $\pm$ 0.098 \\
58669.027 & 20.046 $\pm$ 0.097 & 19.295 $\pm$ 0.075 & 18.975 $\pm$ 0.064 & 19.034 $\pm$ 0.098 \\
58691.94 & 20.521 $\pm$ 0.17 & 19.918 $\pm$ 0.155 & 19.564 $\pm$ 0.144 & 19.657 $\pm$ 0.162 \\
58711.911 & -- & 20.113 $\pm$ 0.269 & 20.018 $\pm$ 0.217 & 19.823 $\pm$ 0.187 \\
58714.563 & 20.724 $\pm$ 0.353 & 20.166 $\pm$ 0.199 & 19.659 $\pm$ 0.162 & 20.229 $\pm$ 0.27 \\
\hline